\documentclass[10pt]{article}

\usepackage{epsfig}
\usepackage{epstopdf}
\usepackage{graphics}
\usepackage{graphicx}
\usepackage{dsfont}
\usepackage{units}
\usepackage{amsmath}
\usepackage{amssymb}
\usepackage{adjustbox}
\usepackage{multirow}
\usepackage{hyperref}
\usepackage{hhline}

\graphicspath{{./eps/}}

\usepackage{cite}

\usepackage{color} 

\usepackage{setspace} 
\usepackage[labelfont=bf,labelsep=period,justification=raggedright]{caption}

\makeatletter
\renewcommand{\@biblabel}[1]{\quad#1.}
\makeatother

\date{}

\newcommand{\hext}{h^{\text{ext}}}
\newcommand{\hextvec}{\vec{h}^{\text{ext}}}
\newcommand{\mninf}{m^n_{\infty}}
\newcommand{\mcinf}{m^c_{\infty}}
\newcommand{\mainf}{m^a_{\infty}}
\newcommand{\ncrit}{n_{\text{crit}}}
\newcommand{\ecrit}{e_{\text{crit}}}
\newcommand{\eopt}{e_{\text{opt}}}

\begin{document}

\begin{flushleft}
{\Large
\textbf{Control of asymmetric Hopfield networks and application to cancer attractors}
}
\\
Anthony Szedlak$^{1,\ast}$, 
Giovanni Paternostro$^{2,3}$, 
Carlo Piermarocchi$^{1,3}$
\\
\bf{1} Department of Physics and Astronomy, Michigan State University, East Lansing MI 48824
\\
\bf{2} Sanford-Burnham Medical Research Institute, La Jolla, CA 92037
\\
\bf{3} Salgomed Inc., Del Mar, CA 92014 USA
\\
$\ast$ E-mail: Corresponding szedlak1@msu.edu
\end{flushleft}


\section*{Abstract}

The asymmetric Hopfield model is used to simulate signaling dynamics in gene regulatory networks. The model allows for a direct mapping of a gene expression pattern into attractor states. We analyze different control strategies aimed at disrupting attractor patterns using selective local fields representing therapeutic interventions. The control strategies are based on the identification of signaling $bottlenecks$, which are single nodes or strongly connected clusters of nodes that have a large impact on the signaling. We provide a theorem with bounds on the minimum number of nodes that guarantee control of bottlenecks consisting of strongly connected components. The control strategies are applied to the identification of sets of proteins that, when inhibited, selectively disrupt the signaling of cancer cells while preserving the signaling of normal cells. We use an experimentally validated non-specific and an algorithmically-assembled specific B cell gene regulatory network reconstructed from gene expression data to model cancer signaling in lung and B cells, respectively. This model could help in the rational design of novel robust therapeutic interventions based on our increasing knowledge of complex gene signaling networks.



\section*{Introduction}

The vision behind systems biology is that complex interactions and emergent properties determine the behavior of biological systems. Many theoretical tools developed in the framework of spin glass models are well suited to describe emergent properties, and their application to large biological networks represents an approach that goes beyond pinpointing the behavior of a few genes or metabolites in a pathway. The Hopfield model~\cite{hopfield1982neural} is a spin glass model that was introduced to describe neural networks, and that is solvable using mean field theory~\cite{amit1985spin}. The asymmetric case, in which the interaction between the spins can be seen as directed, can also be exacty solved in some limits~\cite{derrida1987exactly}. The model belongs to the class of attractor neural networks, in which the spins evolve towards stored attractor patterns, and it has been used to model biological processes of high current interest, such as the reprogramming of pluripotent stem cells \cite{2012arXiv1211.3133L}. Moreover, it has been suggested that a biological system in a chronic or therapy-resistant disease state can be seen as a network that has become trapped in a pathological Hopfield attractor~\cite{10.1371/journal.pone.0014413}. A similar class of models is represented by Random Boolean Networks~\cite{aldana2003boolean}, which were proposed by Kauffman to describe gene regulation and expression states in cells~\cite{Kauffman1969437}. Differences and similarities between the Kauffman-type and Hopfield-type random networks have been studied for many years~\cite{PhysRevE.87.022814,rohlf2009self,0305-4470-21-11-009,Kürten1988157}.

In this paper, we consider an asymmetric Hopfield model built from real (even if incomplete~\cite{de2010advantages,hartemink2005reverse}) cellular networks, and we map the spin attractor states to gene expression data from normal and cancer cells.  We will focus on the question of \emph{controling of a network's final state (after a transient period)} using external local fields representing therapeutic interventions. To a major extent, the final determinant of cellular phenotype is the expression and activity pattern of all proteins within the cell, which is related to levels of mRNA transcripts. Microarrays measure genome-wide levels of mRNA expression that therefore can be considered a rough snapshot of the “state” of the cell. This state is relatively stable, reproducible, unique to cell types, and can differentiate cancer cells from normal cells, as well as differentiate between different types of cancer~\cite{bullinger2004use,eppert2011stem}. In fact, there is evidence that attractors exist in gene expression states, and that these attractors can be reached by different trajectories rather than only by a single transcriptional program~\cite{ PhysRevLett.94.128701}.  While the dynamical attractors paradigm has been originally proposed in the context of cellular developement, the similarity between cellular {\it ontogenesis}, i.e. the developement of different cell types, and {\it oncogenesis}, i.e. the process under which normal cells are transformed into cancer cells, has been recently emphasized~\cite{Huang2009869}.  The main hypothesis of this paper is that cancer robustness is rooted in the dynamical robustness of signaling in an underlying cellular network. If the cancerous state of rapid, uncontrolled growth is an attractor state of the system~\cite{ao2008cancer}, a goal of modeling therapeutic control could be to design complex therapeutic interventions based on drug combinations~\cite{feala2010systems} that “push” the cell out of the cancer attractor basin~\cite{creixell2012navigating}.

Many authors have discussed the control of biological signaling networks using complex external perturbations. Calzolari and coworkers considered the effect of complex external signals on apoptosis signaling~\cite{calzolari2007selective}. Agoston and coworkers~\cite{agoston2005multiple} suggested that perturbing a complex biological network with partial inhibition of many targets could be more effective than the complete inhibition of a single target, and explicitly discussed the implications for multi-drug therapies~\cite{csermely2005efficiency}. In the traditional approach to control theory~\cite{sontag1998mathematical}, the control of a dynamical system consists in finding the specific input temporal sequence required to drive the system to a desired output. This approach has been discussed in the context of Kauffmann Boolean networks~\cite{akutsu2007control} and their attractor states~\cite{choudhary2006intervention}. Several studies have focused on the intrinsic global properties of control and hierarchical organization in biological networks~\cite{feala2012statistical,bhardwaj2010analysis}. A recent study has focused on the minimum number of nodes that needs to be addressed to achieve the complete control of a network~\cite{liu2011controllability}. This study used a linear control framework, a matching algorithm~\cite{plummer1986matching} to find the minimum number of controllers, and a replica method to provide an analytic formulation consistent with the numerical study. Finally, Cornelius {\it et al.}~\cite{cornelius2013realistic} discussed how nonlinearity in network signaling allows reprogrammig a system to a desired attractor state even in the presence of contraints in the nodes that can be accessed by  external control. This novel concept was explicitly applied to a T-cell survival signaling network to identify potential drug targets in T-LGL leukemia. The approach in the present paper is based on nonlinear signaling rules and takes advantage of some useful properties of the Hopfield formulation.  In particular, by considering two attractor states we will show that the network separates into two types of domains which do not interact with each other. Moreover, the Hopfield framework allows for a direct mapping of a gene expression pattern into an attractor state of the signaling dynamics, facilitating the integration of genomic data in the modeling.

The paper is structured as follows. In Mathematical Model we summarize the model and review some of its key properties. Control Strategies describes general strategies aiming at selectively disrupting the signaling only in cells that are near a cancer attractor state. The strategies we have investigated use the concept of $bottlenecks$, which identify single nodes or strongly connected clusters of nodes that have a large impact on the signaling. In this section we also provide a theorem with bounds on the minimum number of nodes that guarantee control of a bottleneck consisting of a strongly connected component. This theorem is useful for practical applications since it helps to establish whether an exhaustive search for such minimal set of nodes is practical. In Cancer Signaling we apply the methods from Control Strategies to lung and B cell cancers. We use two different networks for this analysis. The first is an experimentally validated and non-specific network (that is, the observed interactions are compiled from many experiments conducted on heterogeneous cell types) obtained from a kinase interactome and phospho-protein database~\cite{Yang15082008} combined with a database of interactions between transcription factors and their target genes~\cite{Matys01012003}. The second network is cell-specific and was obtained using network reconstruction algorithms and transcriptional and post-translational data from mature human B cells~\cite{lefebvre2010human}. The algorithmically reconstructed network is significantly more dense than the experimental one, and the same control strategies produce different results in the two cases. Finally, we close with Conclusions.  






\section*{Methods}


\subsection*{Mathematical Model}
\label{sec:model}




We define the adjacency matrix of a network $G$ composed of $N$ nodes as
\begin{equation}
A_{ij}= \left\{
        \begin{array}{ll}
            1 & \text{if }j \rightarrow i \\
            0 & \text{otherwise}
        \end{array}
    \right. \text{ ,}
\end{equation}
\noindent where $j \rightarrow i$ denotes a directed edge from node $j$ to node $i$. The set of nodes in the network $G$ is indicated by $V(G)$ and the set of directed edges is indicated by $E(G)=\left\{(j,i) : j \to i \right\}$. (See Table \ref{table:symbols} for a list of mathematical symbols used in the text.) The spin of node $i$ at time $t$ is 
$\sigma_i(t)=\pm1$, and indicates an expresssed $(+1)$ or not expressed $(-1)$ gene. We encode an arbitrary attractor state  $\vec{\xi}=(\xi_1, \xi_2,...,\xi_N)$ with 
$\xi_i=\pm 1$ by defining the coupling matrix~\cite{hopfield1982neural}
\begin{equation}
J_{ij} = A_{ij}\xi_i\xi_j \text{ .}
\end{equation}
The total field at node $i$ is then 
$
h_i=h^{\text{ext}}_i+\sum_j J_{ij} \sigma_j \text{ ,} \label{eq:h}
$
where $h^{\text{ext}}_i$ is the external field applied to node $i$, which will be discussed below. 
The discrete-time update scheme is defined as
\begin{equation}
\sigma_i(t + \Delta t) = \left\{
        \begin{array}{ll}
            +1 & \text{with prob. } (1+\text{exp}[-h_i(t)/T])^{-1} \\
            -1 & \text{with prob. } (1+\text{exp}[+h_i(t)/T])^{-1}
        \end{array}
    \right. 
\label{eq:update}
\end{equation}
where $T\geq0$ is an effective temperature. For the remainder of the paper, we  consider the case of $T=0$ so that $\sigma_i=\text{sign}(h_i)$, and the spin is chosen randomly from $\pm1$ if $h_i=0$. For convenience, we take $t \in \mathds{Z}$ and $\Delta t=1$. 
Nodes can be updated synchronously, and synchronous updating can lead to limit cycles~\cite{rohlf2009self}.  Nodes can also be updated separately and in random order (anynchronous updating), which does not result in limit cycles. All results presented in this paper use the synchronous update scheme. 

\emph{Source nodes} (nodes with zero indegree) are fixed to their initial states by a small external field so that $\sigma_q(t)=\sigma_q(0)$ for all $q \in Q$, where $Q$ is the set of source nodes. However, the source nodes flip if directly targeted by an external field. Biologically, genes at the ``top'' of a network are assumed to be controlled by elements outside of the network.




In application, two attractors are needed. Define these states as $\vec{\xi}^n$ and $\vec{\xi}^c$, the \emph{normal state} and \emph{cancer state}, respectively. The magnetization along attractor state $a$ is
\begin{equation}
m^a(t) = \frac{1}{N} \sum_{i=1}^N \sigma_i(t) \xi_i^a \text{ .}
\end{equation}
Note that if $m^a(t)=\pm1$, $\vec{\sigma}(t)=\pm\vec{\xi}^a$. We also define the steady state magnetization along state $a$ as
\begin{equation}
m^a_{\infty} = \lim_{\tau \to \infty} \frac{1}{\tau} \sum_{t=1}^{\tau} m^a(t) \text{ .}
\end{equation}


There are two ways to model normal and cancer cells. One way is to simply define a different coupling matrix for each attractor state $a$,
\begin{equation}
J^a_{ij} = A_{ij} \xi_i^a \xi_j^a \text{ .} \label{eq:coup_mat_p_eq_1}
\end{equation}

\noindent Alternatively, both attractor states can be encoded in the same coupling matrix,
\begin{equation}
J_{ij} = A_{ij}(\xi_i^n\xi_j^n + \xi_i^c\xi_j^c) \text{ .} \label{eq:coup_mat_p_eq_2}
\end{equation}

\noindent Systems using Eqs. \ref{eq:coup_mat_p_eq_1} and \ref{eq:coup_mat_p_eq_2} will be referred to as the one attractor state ($p=1$) and two attractor state ($p=2$) systems, respectively.
Eqs. \ref{eq:coup_mat_p_eq_1} and \ref{eq:coup_mat_p_eq_2} are particular cases of the  general Hopfield form~\cite{hopfield1982neural}
\begin{equation}
J_{ij} = A_{ij} \sum_{k=1}^p \xi^k_i \xi^k_j \text{ ,}
\end{equation}
where $p$ is the number of attractor states, often taken to be large. An interesting property emerges when $p=2$, however. Consider a simple network composed of two nodes, with only one edge $1\to2$ with attractor states $\vec{\xi}^n$ and $\vec{\xi}^c$, and $T=0$. The only nonzero entry of the matrix $J_{ij}$ is
\begin{equation}
J_{21} = \xi^n_2 \xi^n_1 + \xi^c_2 \xi^c_1 \text{ .}
\end{equation}
Note that if $\vec{\xi}^n=\pm \vec{\xi}^c$, $J_{21}=2\xi^n_2 \xi^n_1$. In either case, by Eq. \ref{eq:update} we have 
\begin{equation}
\sigma_2(t + 1) = \left\{
        \begin{array}{ll}
            +\xi^n_2 & \text{if } \sigma_1(t)=+\xi^n_1 \\
            -\xi^n_2 & \text{if } \sigma_1(t)=-\xi^n_1
        \end{array}
    \right. \text{ ,}
\end{equation}
\noindent that is, the spin of node 2 at a given time step will be driven to match the attractor state of node 1 at the previous time step. However, if $\xi^n_1=\pm \xi^c_1$ and $\xi^n_2=\mp \xi^c_2$, $J_{21}=0$. This gives
\begin{equation}
\sigma_2(t)= \left\{
        \begin{array}{ll}
            +1 & \text{with probability } \nicefrac{1}{2} \\
            -1 & \text{with probability } \nicefrac{1}{2}
        \end{array}
    \right.
\end{equation}
\noindent In this case, node 2 receives no input from node 1. Nodes 1 and 2 have become effectively disconnected.

\begin{figure}[ht]
\begin{center}
\includegraphics[width=4in]{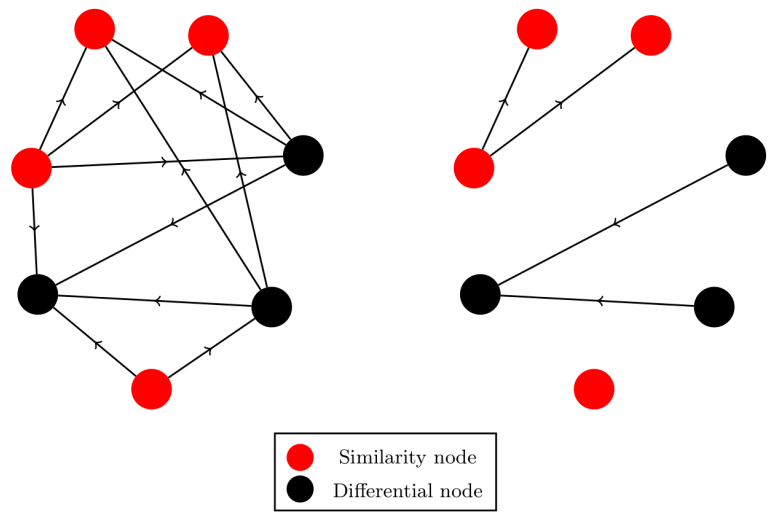}
\end{center}
\caption{{\bf Network segregation for two attractor states ($p=2$).} Every edge that connects a similarity node to a differential node or a differential node to a similarity node transmits no signal. This means that the signaling in the right network shown above is identical to that of the left network. Because the goal is to leave normal cells unaltered while damaging cancer cells as much as possible, all similarity nodes can be safely ignored, and searches and simulations only need to be done on the differential subnetwork.}
\label{fig:sim_diff_sep}
\end{figure}

This motivates new designations for node types. We define \emph{similarity nodes} as nodes with $\xi_i^n=\xi_i^c$, and \emph{differential nodes} as nodes with $\xi_i^n=-\xi_i^c$. We also define the set of similarity nodes $S=\left\{i:\xi_i^n=\xi_i^c\right\}$ and the set of differential nodes $D=\left\{i:\xi_i^n=-\xi_i^c\right\}$. Connections between two similarity nodes or two differential nodes remain in the network, whereas connections that link nodes of different types transmit no signals. The effective deletion of edges between nodes means that the original network fully separates into two subnetworks: one composed entirely of similarity nodes (the \emph{similarity network}) and another composed entirely of differential nodes (the \emph{differential network}), each of which can be composed of one or more separate weakly connected components (see Fig. \ref{fig:sim_diff_sep}). With this separation, new source nodes (\emph{effective sources}) can be exposed in both the similarity and differential networks. For the remainder of this article, $Q$ is the set of both source and effective source nodes in a given network.

\subsection*{Control Strategies}
\label{sec:strategies}

The strategies presented below focus on selecting the best single nodes or small clusters of nodes to control, ranked by how much they individually change $m^a_{\infty}$. In application, however, controlling many nodes is necessary to achieve a sufficiently changed $m^a_{\infty}$. The effects of controlling a set of nodes can be more than the sum of the effects of controlling individual nodes, and predicting the truly optimal set of nodes to target is computationally difficult. Here, we discuss heuristic strategies for controlling large networks where the combinatorial approach is impractical. 

For both $p=1$ and $p=2$, simulating a cancer cell means that $\vec{\sigma}(0)=+\vec{\xi}^c$, and likewise for normal cells. Although the normal and cancer states are mathematically interchangeable, biologically we seek to decrease $\mcinf$ as much as possible while leaving $\mninf\approx+1$. 
By ``network control''  we thus mean driving the system away from its initial state of $\vec{\sigma}(0)=\vec{\xi}^c$ with $\hextvec$.
Controlling individual nodes is achieved by applying a strong field (stronger than the magnitude of the field due to the node's upstream neighbors) to a set of targeted nodes $T$ so that
\begin{equation}
\hext_{\tau}= \left\{
        \begin{array}{ll}
            \lim_{(u \to \infty)} -u\xi^c_{\tau} & \tau \in T \\
            0 & \text{else}
        \end{array}
    \right. \text{ .}
\end{equation}
This ensures that the drug field can always overcome the field from neighboring nodes.

In application, similarity nodes are never deliberately directly targeted, since changing their state would adversely affect both normal and cancer cells. Roughly $70\%$ of the nodes in the networks surveyed are similarity nodes, so the search space is reduced. For $p=2$, the effective edge deletion means that only the differential network in cancer cells needs to be simulated to determine the effectiveness of $\hextvec$. For $p=1$, however, there may be some similarity nodes that receive signals from upstream differential nodes. In this case, the full effect of $\hextvec$ can be determined only by simulating all differential nodes as well as any similarity nodes downstream of differential nodes. All following discussion assumes that all nodes examined are differential, and therefore targetable, for both $p=1$ and $p=2$. The existence of similarity nodes for $p=1$ only limits the set of targetable nodes.

\subsubsection*{Directed acyclic networks}
Full control of a directed acyclic network is achieved by forcing $\sigma_q=-\xi_q^c$ for all $q\in Q$. This guarantees $\mcinf=-1$. Suppose that nodes $q \in Q$ in an acyclic network have always been fixed away from the cancer state, that is, $\sigma_q(t \to -\infty) = -\xi^c_q$. For any node $i$ to have $\sigma_i(t)=\xi^n_i$, it is sufficient to have either $i \in Q$ or $\sigma_j(t-1)=\xi^n_j$ for all $j \to i$, $i \notin Q$. Because there are no cycles present, all upstream paths of sufficent length terminate at a source. Because the spin of all nodes $q \in Q$ point away from the cancer attractor state, all nodes downstream must also point away from the cancer attractor state. Thus, for acyclic networks, forcing $\sigma_q=-\xi_q^c$ guarantees $\mcinf=-1$. The complications that arise from cycles are discussed in the next subsubsection. However, controlling nodes in $Q$ may not be the most efficient way to push the system away from the cancer basin of attraction and, depending on the control limitations, it may not be possible. If minimizing the number of controllers is required, searching for the most important bottlenecks is a better strategy. 

Consider a directed network $G$ and an initially identical copy, $G^{\prime} = G$. If removing node $i$ (and all connections to and from $i$) from $G^{\prime}$ decreases the indegree of at least one node $j\in V(G^{\prime})$, $j \neq i$, to less than half of its indegree in network $G$, $\{i\}$ is a \emph{size 1 bottleneck}.
The \emph{bottleneck control set} of bottleneck $\{i\}$, $L(i)$, is defined algorithmically as follows: 
(1) Begin a set $L(i)$ with the current bottleneck $i$ so that $L=\{ i \}$; 
(2) Remove bottleneck $\{i\}$ from network $G^{\prime}$;
(3) Append $L(i)$ with all nodes $j$ with current indegree that is less than half of that from the original network $G$;
(4) Remove all nodes $j$ from the network $G^{\prime}$. If additional nodes in $G^{\prime}$ have their indegree reduced to below half of their indegree in $G$, go to step 3. Otherwise, stop.
The \emph{impact of the bottleneck i}, $I(i)$, is defined as
\begin{equation}
I(i)=|L(i)|    \text{ ,}
\end{equation}
where $|X|$ is the cardinality of the set $X$. The impact of a bottleneck is the minimum number of nodes that are guaranteed to switch away from the cancer state when the bottleneck is forced away from the cancer state. 

The impact is used to rank the size 1 bottlenecks by importance, with the most important as those with the largest impact. In application, when searching for nodes to control, any size 1 bottleneck $\{i\}$ that appears in the bottleneck control set of a different size 1 bottleneck $\{j\}$ can be ignored, since fixing $j$ to the normal state fixes $i$ to the normal state as well. Note that the definition given above in terms of $G$ and $G^\prime$ avoids miscounting in the impact of a bottleneck.

%
%
%
%

\begin{figure}[!ht]
\begin{center}
\includegraphics[width=4in]{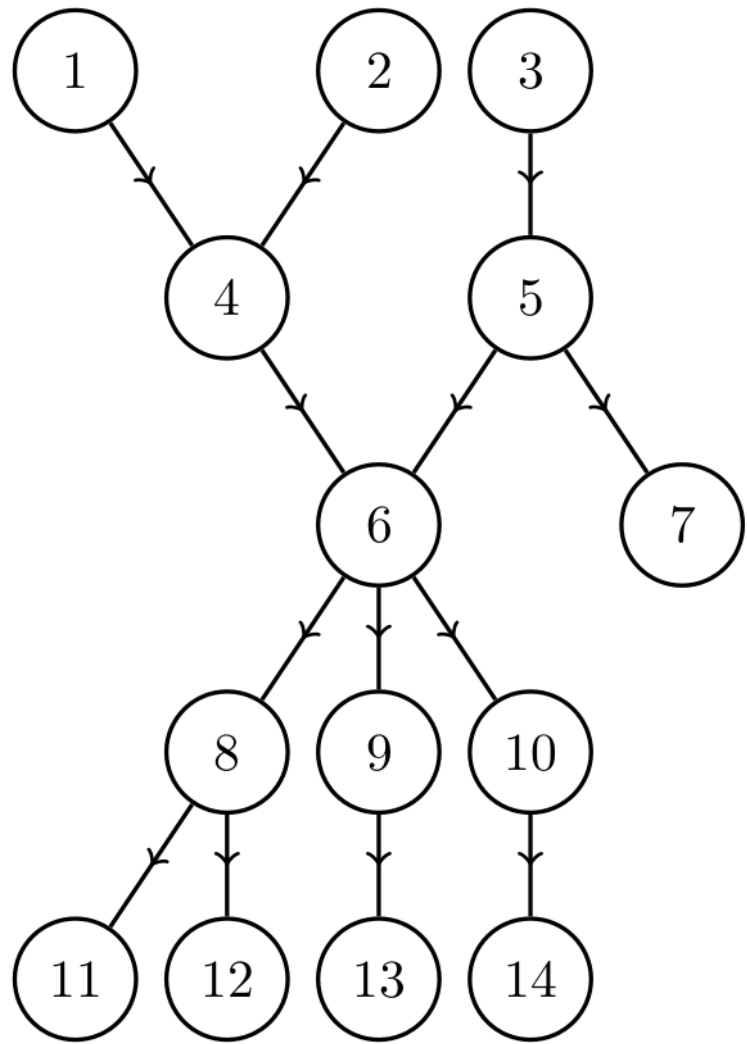}
\end{center}
\caption{{\bf A directed acyclic network.} Controlling all three source nodes (nodes 1, 2 and 3) guarantees full control of the network, but are ineffective when targeted individually. The best single node to control in this network is node 6 because it directly controls all downstream nodes.}
\label{fig:acyc}
\end{figure}

The network in Fig. \ref{fig:acyc}, for example, has three sources (nodes 1, 2 and 3), but one  important bottleneck (node 6). If full damage, i.e. $\mcinf=-1$, is required, then control of all source nodes is necessary. If minimizing the number of directly targeted nodes is important and $\mcinf > -1$ can be tolerated, then control of the bottleneck node 6 is a better choice. 

\subsubsection*{Directed cycle-rich networks}

Not all networks can be fully controlled at $T=0$ by controlling the source nodes, however. If there is a cycle present, paths of infinite length exist and the final state of the system may depend on the initial state, causing parts of the network to be hysteretic. Controlling only sources in a general directed network thus does not guarantee $\mcinf=-1$ unless the system begins with $\sigma_i=-\xi^c_i$.

\begin{figure}[!ht]
\begin{center}
\includegraphics[width=4in]{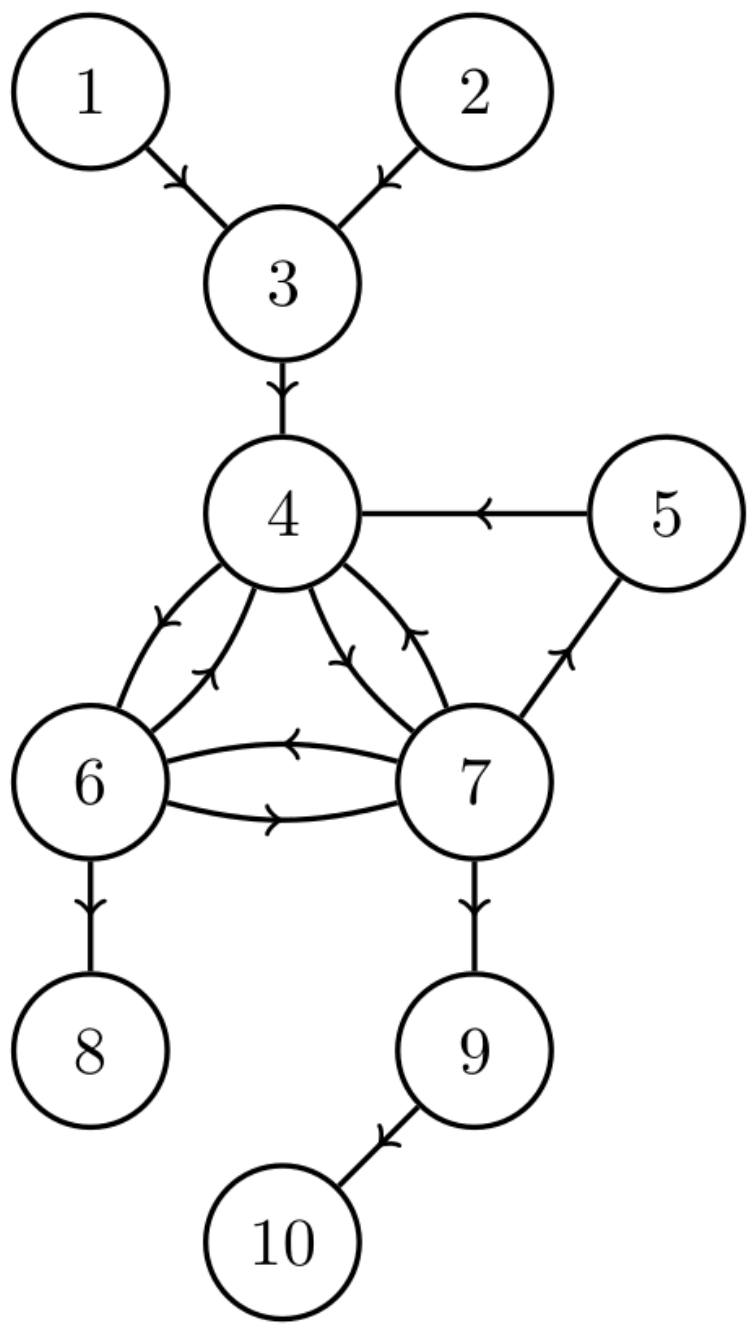}
\end{center}
\caption{{\bf A network in which nodes 4, 5, 6 and 7 compose a single cycle cluster.} The high connectivity of node 4 prevents any changes made to the spin of nodes 1-3 from propagating downstream. The only way to indirectly control nodes 8-10 is to target nodes inside of the cycle cluster. Targeting node 4, 6 or 7 will cause the entire cycle cluster to flip away from its initial state, guaranteeing control of nodes 4-10 (see Fig. \ref{fig:cyc_plot}).}
\label{fig:cyc}
\end{figure}

\begin{figure}[!ht]
\begin{center}
\includegraphics[width=0.5\textwidth]{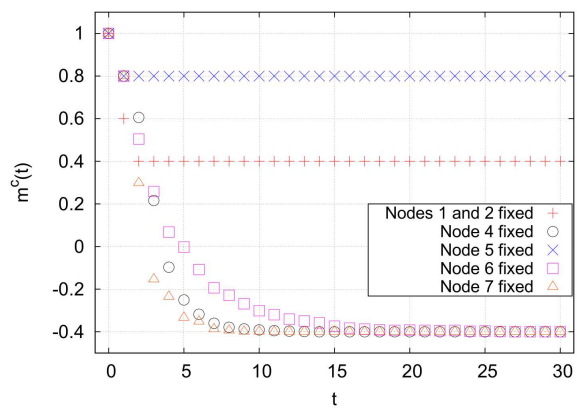}
\end{center}
\caption{{\bf Cancer magnetization from targeting various nodes in the network shown in Fig. \ref{fig:cyc}, averaged over 10,000 runs.} The averaging removes fluctuations due to the random flipping of nodes with $h_i=0$.  Targeting node 7 results in the quickest stabilization, but targeting any one of nodes 4, 6 or 7 results in the same final magnetization.}
\label{fig:cyc_plot}
\end{figure}

Define a \emph{cycle cluster}, $C$, as a strongly connected subnetwork of a network $G$. The network in Fig. \ref{fig:cyc}, for example, has one cycle cluster with nodes $V(C)=\left\{4,5,6,7\right\}$. If the network begins with $\vec{\sigma}(0) = \vec{\xi}^c$, forcing both source nodes away from the cancer state does nothing to the nodes downsteam of node 3 (see Fig. \ref{fig:cyc_plot}). This is because the indegree $\text{deg}^-(4)=4$, and a majority of the nodes connecting to node 4 are in the cancer attractor state. At $T=0$, cycle clusters with high connectivity tend to block incoming signals from outside of the cluster, resulting in an insurmountable activation barrier.

The most effective single node to control in this network is any one of nodes 4, 6 or 7. Forcing any of these away from the cancer attractor state will eventually cause the entire cycle cluster to flip away from the cancer state, and all nodes downstream will flip as well, as shown in Fig. \ref{fig:cyc_plot}. The cycle cluster here acts as a sort of large, hysteretic bottleneck. We now generalize the concept of bottlenecks.

Define a \emph{size $k$ bottleneck} in a network $G$ to be a cycle cluster $B$ with $|V(B)|=k$ which, when removed from $G$, reduces the indegree of at least one node $j\in V(G)$, $j \not\in V(B)$ to less than half of its original indegree. Other than now using the set of nodes $V(B)$ rather than a single node set, the above algorithm for finding the bottleneck control set remains unchanged. In Fig. \ref{fig:cyc}, for instance, $V(B)=\left\{ 4,5,6,7 \right\}$, $k=4$, $L(B)=\left\{ 4,5,6,7,8,9,10 \right\}$, and $I(B)=7$. With this more general definition, we note that controlling any size $k$ bottleneck $B$ guarantees control of all size 1 bottlenecks $B^\prime$ in the control set of $B$ for all $k\geq1$.

For any bottleneck $B$ of size $k\geq1$ in a network $G$, define the \emph{set of critical nodes}, $Z(B,G)$, as the set of nodes $Z(B,G) \subseteq V(B)$ of minimum cardinality that, when controlled, guarantees full control of all nodes $i\in V(B)$ after a transient period. Also define the \emph{critical number of nodes} as $n_{\text{crit}}(B,G)=|Z(B,G)|$. Thus, for the network in Fig. \ref{fig:cyc}, $Z(B,G)=\{4\}$, $\{6\}$, or $\{7\}$, and $n_{\text{crit}}(B,G)=1$.

\begin{figure}[!ht]
\begin{center}
\includegraphics[width=4in]{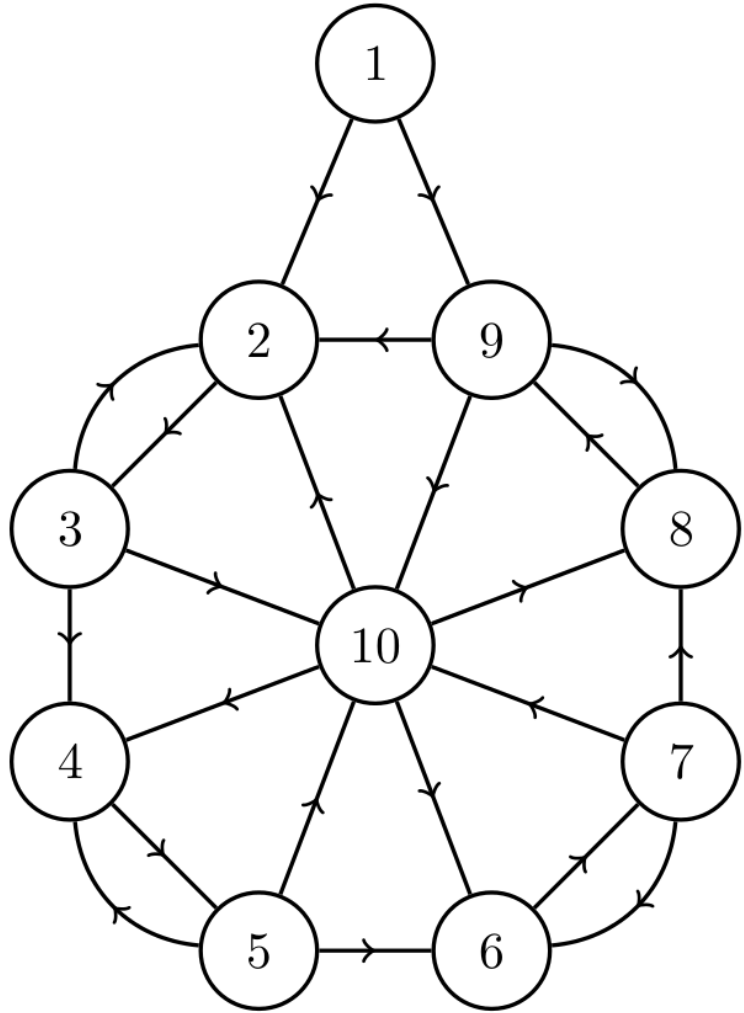}
\end{center}
\caption{{\bf A network with a cycle cluster $C$, composed of nodes 2-10, that cannot be controlled at $T=0$ by controlling any single node.} Here, the set of externally influenced nodes is $R(C,G)=\{2,9\}$, the set of intruder connections is $W(C,G)=\{(1,2),(1,9)\}$, the reduced set of critical nodes is $Z_{\text{red}}(C,G)=\{9,10\}$, the minimum indegree is $\mu=1$ and the number of nodes in the cycle cluster is $N=9$. By Eq. \ref{eq:n_crit_bounds}, this gives the bounds of the critical number of nodes to be $1\leq\ncrit\leq6$.}
\label{fig:contrived}
\end{figure}

In general, however, more than one node in a cycle cluster may need to be targeted to control the entire cycle cluster. Fig. \ref{fig:contrived} shows a cycle cluster (composed of nodes 2-10) that cannot be controlled by targeting any single node. The precise value of $\ncrit$ for a given cycle cluster $C$ depends on its topology as well as the edges connecting nodes from outside of $C$ to the nodes inside of $C$, and finding $Z(C,G)$ can be difficult. We present a theorem that puts bounds on $\ncrit$ to help determine whether a search for $Z(C,G)$ is practical.

\emph{Theorem:} Suppose a network $G$ contains a cycle cluster $C$. Define the \emph{set of externally influenced nodes}
\begin{equation}
R(C,G)=\{i\in V(C) : j\in V(G\setminus C), (j,i)\in E(G)\}~,
\end{equation}
the \emph{set of intruder connections}
\begin{equation}
W(C,G)=\{(j,i) \in E(G) : i \in V(C), j \in V(G\setminus C)\} \text{~,}
\end{equation}
and the \emph{reduced set of critical nodes}
\begin{equation}
Z_{\text{red}}(C,G)=Z(C,G \setminus W) \text{~.}
\label{eq:z_red}
\end{equation}
If $N=|V(C)|$ and
\begin{equation}
\mu \equiv \smash{\min_{i \in V(C)}} \text{deg}^-(i) \text{ ,}
\label{eq:min_degree}
\end{equation}
where $\text{deg}^-(i)$ is computed ignoring intruder connections, then
\begin{equation}
\left \lceil \frac{\mu}{2} \right \rceil \leq n_{\text{crit}}(C,G) \leq \zeta \text{ ,}
\label{eq:n_crit_bounds}
\end{equation}
where
\begin{equation}
\zeta \equiv \text{min} \left ( \left \lceil \frac{N}{2} \right \rceil + |R(C,G) \setminus Z_{\text{red}}(C,G)| , N \right ) \text{ .}
\end{equation}

\emph{Proof:} First, prove the lower limit of Eq. \ref{eq:n_crit_bounds}. Let $C$ be a cycle cluster in a network $G$ with $R(C,G)=\{\varnothing\}$. (A cycle cluster in a network with $|R(C,G)|>0$ will have the same or higher activation barrier for any node in the cluster than the same cycle cluster in a network with $R=\{\varnothing\}$. Since we are examining the lower limit of Eq. \ref{eq:n_crit_bounds}, we consider the case with the lowest activation barrier. Any externally influenced nodes cause $n_{\text{crit}}$ to either increase or remain the same.) For any node $i$ to be able to flip away from the cancer state (although not necessarily remain there), we must have that $h_i = -a \xi^c_i$ for $a \geq 0$, meaning that at least half of the nodes upstream of $i$ must point away from the cancer state. The node $i$ requiring the smallest number of upstream nodes to be in the normal state is the node that satisfies $\text{deg}^-(i)=\mu$. Controlling less than $\mu/2$ nodes will leave all uncontrolled nodes with a field in the cancer direction, and no more flips will occur. Thus,
\begin{equation}
n_{\text{crit}} \geq \left\lceil \frac{\mu}{2} \right\rceil \text{ .}
\label{eq:n_crit_lower}
\end{equation}

For the upper limit of Eq. \ref{eq:n_crit_bounds}, consider a complete \emph{clique} on $N$ nodes, $C=K_N$ (that is, $A_{ij}=1$ for all $i,j \in V(K_N)$, including self loops) in a network $G$. First, let there be no connections to any nodes in $C$ from outside of $C$ so that $R(C,G)=\{\varnothing\}$. For odd $N$, forcing $(N+1)/2$ nodes away from the cancer state will result in the field
\begin{equation}
\sum_j J_{ij} \sigma_j = \left(\frac{N-1}{2} - \frac{N+1}{2}\right)\xi^c_i = -\xi^c_i
\end{equation}

\noindent for all nodes $i$. After one time step, all nodes will flip away from the cancer state. For even $N$, forcing $N/2$ nodes away from the cancer state will result in the field
\begin{equation}
\sum_j J_{ij} \sigma_j = \left(\frac{N}{2} - \frac{N}{2}\right)\xi^c_i = 0
\end{equation}

\noindent for all nodes $i$. At the next time step, the unfixed nodes will pick randomly between the normal and cancer state. If at least one of these nodes makes the transition away from the cancer state, the field at all other nodes will point away from the cancer direction. The system will then require one more time step to completely settle to $\sigma_i=-\xi^c_i$. Thus, we have that for $C=K_N$ in a network $G$ with $R(C,G)=\{ \varnothing \}$,
\begin{equation}
n_{\text{crit}}(K_N,G) = \left \lceil \frac{N}{2} \right \rceil \text{ .}
\end{equation}

\noindent $K_N$ with $\sigma_i(0)=\xi^c_i$ gives the largest activation barrier for any cycle cluster on $N$ nodes with $R(C,G)=\{\varnothing\}$ to switch away from the cancer attractor state. A general cycle cluster $C$ with any topology on $N$ nodes with $R(C,G)=\{\varnothing\}$ in a network $G$ will have $\text{deg}^-(i) \leq N$ for all nodes $i$, and so we have the upper bound
\begin{equation}
n_{\text{crit}}(C,G) \leq \left \lceil \frac{N}{2} \right \rceil \text{ ,}
\label{eq:n_crit_special_case}
\end{equation}

\noindent thus proving Eq. \ref{eq:n_crit_bounds} for the special case of $R(C,G)=\{\varnothing\}$.

Now consider a cycle cluster $C$ on $N$ nodes in a network $G$ with $|R(C,G)|\geq0$. Suppose all nodes in $Z_{\text{red}}(C,G)$ are fixed away from the cancer state. By Eq. \ref{eq:n_crit_special_case}, $|Z_{\text{red}}(C,G)| \leq \lceil N/2 \rceil$. For any node $i \in (R(C,G) \cap Z_{\text{red}}(C,G))$, $\sigma_i(t \to \infty)=-\xi^c_i$ is guaranteed because it has already been directly controlled. Any node $i\in (R(C,G)\setminus Z_{\text{red}}(C,G))$ has some incoming connections from nodes $j \notin V(C)$, and these connections could increase the activation barrier enough such that fixing $Z_{\text{red}}(C,G)$ is not enough to guarantee $\sigma_i(t \to \infty)=-\xi^c_i$. To ensure that any node $l\in V(C)$ points away from the cancer state, it is sufficient to fix all nodes $i\in (R(C,G)\setminus Z_{\text{red}}(C,G))$ as well as $Z_{\text{red}}(C,G)$ away from the cancer state. This increases $n_{\text{crit}}$ by at most $|R(C,G)\setminus Z_{\text{red}}(C,G)|$, leaving
\begin{equation}
n_{\text{crit}}(C,G) \leq \left \lceil \frac{N}{2} \right \rceil + |R(C,G)\setminus Z_{\text{red}}(C,G)| \text{ .}
\end{equation}

\noindent $\ncrit$ can never exceed $N$, however, because directly controlling every node results in controlling $C$. We can thus say that
\begin{equation}
n_{\text{crit}}(C,G) \leq \text{min} \left ( \left \lceil \frac{N}{2} \right \rceil + |R(C,G)\setminus Z_{\text{red}}(C,G)|,N \right ) \text{ .}
\label{eq:n_crit_upper}
\end{equation}

\noindent Finally, combining the upper limit in Eq. \ref{eq:n_crit_upper} with the lower limit from Eq. \ref{eq:n_crit_lower} gives Eq. \ref{eq:n_crit_bounds}. $\blacksquare$

There can be more than one $Z_{\text{red}}$ for a given cycle cluster. Note that the tightest constraints on $n_{\text{crit}}$ in Eq. \ref{eq:n_crit_bounds} come from using the $Z_{\text{red}}$ with the largest overlap with $R$. If finding $Z_{\text{red}}$ is too difficult, an overestimate for the upper limit of $n_{\text{crit}}$ can be made by assuming that $R \cap Z_{\text{red}} = \{\varnothing\}$ so that
\begin{equation}
\left \lceil \frac{\mu}{2} \right \rceil \leq n_{\text{crit}}(C,G) \leq \min \left ( \left \lceil \frac{N}{2} \right \rceil + |R(C,G)| \text{ },\text{ } N \right ) \text{ .}
\label{eq:n_crit_bounds_worst_case}
\end{equation}

\begin{figure}[!ht]
\begin{center}
\includegraphics[width=0.5\textwidth]{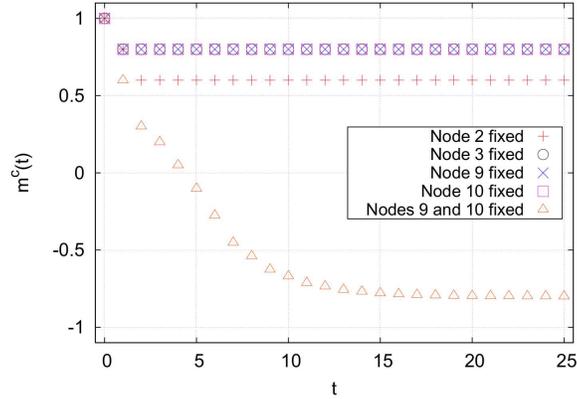}
\end{center}
\caption{{\bf Magnetization for network from Fig. \ref{fig:contrived}, averaged over 10,000 runs.} There is no single node to target that will control the cycle cluster, but fixing nodes 9 and 10 results in full control of the cycle cluster, leaving only node 1 in the cancer state. This means $Z(C,G)=\{9,10\}$ and  $\ncrit=2$.}
\label{fig:contrived_plot}
\end{figure}

The cycle cluster in Fig. \ref{fig:contrived} has $N=9$, $R=\{2,9\}$, $\mu=1$, and one of the reduced sets of critical nodes is $Z_{\text{red}}=\{9,10\}$, so $1 \leq n_{\text{crit}} \leq 6$. It can be shown through an exhaustive search that for this network $n_{\text{crit}}=2$, and the set of critical nodes is $Z=\{9,10\}$ (see Fig. \ref{fig:contrived_plot}). Here, $Z=Z_{\text{red}}$, although this is not always the case. Because the cycle cluster has 9 nodes and $1 \leq n_{\text{crit}} \leq 6$, at most $\sum_{n=1}^6{9 \choose n}=465$ simulations are needed to find at least one solution for $Z(C,G)$. However, the maximum number of simulations required to find $Z(C,G)$ increases exponentially and for larger networks the problem quickly becomes intractable. 

One heuristic strategy for controlling cycle clusters is to look for size $k^{\prime}<|V(C)|$ bottlenecks inside of $C$. Bottlenecks of size $k\gg 1$ and average indegree $\langle \text{deg}^-(B) \rangle \ll k$ can contain high impact size $k^{\prime}$ bottlenecks, where $k^{\prime}<k$.
Size $k\geq1$ bottlenecks need to be compared to find the best set of nodes to target to reduce $\mcinf$. Simply comparing the impact is insufficent because a cycle cluster with a large impact could also have a large $n_{\text{crit}}$, requiring much more effort than its impact merits. Define the \emph{critical efficiency} of a bottleneck $B$ as
\begin{equation}
\ecrit(B)=\frac{I(B)}{n_{\text{crit}}(B,G)} \text{ .}
\end{equation}

\noindent If the critical efficiency of a cycle cluster is much smaller than the impacts of size 1 bottlenecks from outside of the cycle cluster, the the cycle cluster can be safely ignored.

For some cycle clusters, however, not all of the nodes need to be controlled in order for a large portion of the nodes in the cycle cluster's control set to flip. Define the \emph{optimal efficiency} of a bottleneck $B$ as
\begin{equation}
\eopt(B) = \smash{\displaystyle\max_{n=1,2,\ldots}}\left(\frac{I\left(\bigcup_{i=1}^{n} B_i\right)}{n}\right)
\end{equation}

\noindent where $B_i \subseteq V(B)$ are size 1 bottlenecks and $I(B_i)>I(B_{i+1})$ for all $i$. Note that for any size 1 bottleneck $B$, $\eopt(B)=\ecrit(B)=I(B)$. This quantity thus allows bottlenecks with very different properties ($I(B)$, $n_{\text{crit}}(B,G)$, or $|V(B)|$) to be ranked against each other.

All strategies presented above are designed to select the best individual or small group of nodes to target. Significant changes in the biological networks' magnetization require targeting many nodes, however. Brute force searches on the effect of larger combinations of nodes are typically impossible because the required number of simulations scales exponentially with the number of nodes. A crude Monte Carlo search is also numerically expensive, since it is difficult to sample an appreciable portion of the available space. One alternative is to take advantage of the bottlenecks that can be easily found, and rank all size $k\geq1$ bottlenecks $B_i$ in an ordered list $U$ such that
\begin{equation}
U = (B_1,B_2,B_3,\ldots)
\end{equation}

\noindent where
\begin{equation}
\eopt(B_i) \geq \eopt(B_{i+1}) \text{ , } B_i \not\subset L(B_j)
\end{equation}

\noindent for all $B_i,B_j \in U$ and fix the bottlenecks in the list in order. This is called the \emph{efficiency-ranked} strategy. If all size $k>1$ bottlenecks are ignored, it is called the \emph{pure} efficiency-ranked strategy, and if size $k>1$ bottlenecks are included it is called the \emph{mixed} efficiency-ranked strategy. 

An effective polynomial-time algorithm for finding the top $z$ nodes to fix, which we call the \emph{best+1} strategy (equivalent to a greedy algorithm), works as follows:
(1) Begin with a seed set of nodes to fix, $F$;
(2) Test the effect of fixing $F\cup i$ for all allowed nodes $i\notin F$;
(3) $F\gets F\cup i_{\text{best}}$, where $i_{\text{best}}$ is the best node from all $i$ sampled;
(4) If $|F|<z$, go to step (2). Otherwise, stop.
The seed set of nodes could be the single highest impact size 1 bottleneck in the network, or it could be the best set of $n$ nodes (where $n<z$) found from a brute force search.


\section*{Cancer Signaling}
\label{sec:results}

In application to biological systems, we assume that the magnetization of cell type $a$ is related to the \emph{viability} of cell type $a$, that is, the fraction of cells of type $a$ that survives a drug treatment. It is reasonable to assume that the viability of cell type $a$, $v^a(\mainf)$, is a monotonically increasing function of $\mainf$. Because the exact relationship is not known, we analyze the effect of external perturbations in terms of the final magnetizations.

We need to use as few controllers as possible to sufficiently reduce $\mcinf$ while leaving $\mninf\approx+1$. In practical applications, however, one is limited in the set of druggable targets. All classes of drugs are constrained to act only on a specific set of biological components. For example, one class of drugs that is currently under intense research is protein kinase inhibitors~\cite{cohen2002protein}. In this case one has two constraints: the only nodes that can be targeted are those that correspond to kinases, and they can only be inhibited, i.e. turned off. We will use the example of kinase inhibitors to show how control is affected by such types of constraints. In the real systems studied, many differential nodes have only similarity nodes upstream and downstream of them, while the remaining differential nodes form one large cluster. This is not important for $p=1$, but the effective edge deletion for $p=2$ results in many \emph{islets}, which are nodes $i$ with $A_{ij}=A_{ji}=0$ for all $i \neq j$ (self-loops allowed). Controlling islets requires targeting each islet individually. For $p=2$, we concentrate on controlling only the largest weakly connected differential subnetwork. All final magnetizations are normalized by the total number of nodes in the full network, even if the simulations are only conducted on small portion of the network.

The data files for all networks and attractors analyzed below can be found in Supplementary Information.


\subsection*{Lung Cell Network}
The network used to simulate lung cells was built by combining the kinase interactome from PhosphoPOINT~\cite{Yang15082008}  with the transcription factor interactome from TRANSFAC~\cite{Matys01012003}. Both of these are general networks that were constructed by compiling many observed pairwise interactions between components, meaning that if $j\to i$, at least one of the proteins encoded by gene $j$ has been directly observed interacting with gene $i$ in experiments. This bottom-up approach means that some edges may be missing, but those present are reliable. Because of this, the network is sparse ($\sim0.057\%$ complete, see Table \ref{table:general}), resulting in the formation of many islets for $p=2$. Note also that this network presents a clear hierarchical structure, characteristic of biological networks~\cite{ravasz2002hierarchical,girvan2002community}, with many "sink'' nodes~\cite{shen2002network} that are targets of transcription factors and a relatively large cycle cluster originating from the kinase interactome.

It is important to note that this is a non-specific network, whereas real gene regulatory networks can experience a sort of ``rewiring'' for a single cell type under various internal conditions~\cite{gerstein2004topologicalchanges}. In this analysis, we assume that the difference in topology between a normal and a cancer cell's regulatory network is negligible. The methods described here can be applied to more specialized networks for specific cell types and cancer types as these networks become more widely avaliable.

In our signaling model, the IMR-90 cell line~\cite{munoz2011quantitative,muggerud2009data} was used for the normal attractor state, and the two cancer attractor states examined were from the A549 (adenocarcinoma)~\cite{wagner2007death,stinson2011trps1,hussain2009tobacco,muzikar2009repression,sartor2010conceptgen} and NCI-H358 (bronchioalveolar carcinoma)~\cite{wagner2007death,stinson2011trps1} cell lines. Gene expression measurements from all referenced studies for a given cell line were averaged together to create a single attractor. The resulting magnetization curves for A549 and NCI-H358 are very similar, so the following analysis addresses only A549. 
The full network contains 9073 nodes, but only 1175 of them are differential nodes in the IMR-90/A549 model. In the unconstrained $p=1$ case, all 1175 differential nodes are candidates for targeting. Exhaustively searching for the best pair of nodes to control requires investigating 689725 combinations simulated on the full network of 9073 nodes. However, 1094 of the 1175 nodes are sinks (i.e. nodes $i$ with outdegree $\text{deg}^+(i)=0$, ignoring self loops) and therefore have $I(i)=\eopt(i)=1$, which can be safely ignored. The search space is thus reduced to 81 nodes, and finding even the best triplet of nodes exhaustively is possible. Including constraints, only 31 nodes are differential kinases with $\xi^c_i=+1$. This reduces the search space at the cost of increasing the minimum achievable $\mcinf$.

There is one important cycle cluster in the full network, and it is composed of 401 nodes. This cycle cluster has an impact of 7948 for $p=1$, giving a critical efficiency of at least $\sim19.8$, and $1 \leq n_{\text{crit}} \leq 401$ by Eq. \ref{eq:n_crit_bounds_worst_case}.
The optimal efficiency for this cycle cluster is $\eopt = 29$, but this is achieved for fixing the first bottleneck in the cluster. Additionally, this node is the highest impact size 1 bottleneck in the full network, and so the mixed efficiency-ranked results are identical to the pure efficiency-ranked results for the unconstrained $p=1$ lung network. The mixed efficiency-ranked strategy was thus ignored in this case. 

\begin{figure}[!ht]
\begin{center}
\includegraphics[width=0.5\textwidth]{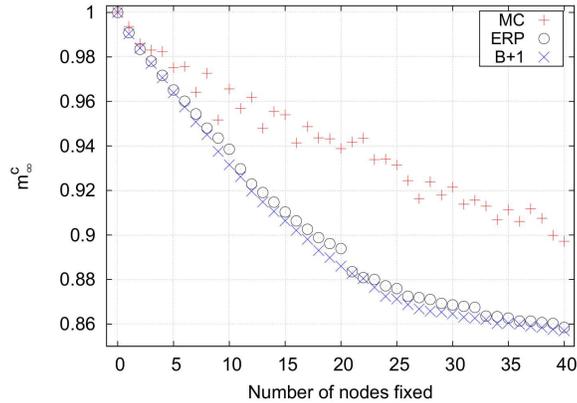}
\end{center}
\caption{{\bf Final cancer magnetizations for an unconstrained search on the lung cell network using $p=1$.} The efficiency-ranked strategy outperforms the relatively expensive Monte Carlo strategy. The best+1 strategy works best, although it requires the largest computational time. Note that the mixed efficiency-ranked curve is not shown because it is identical to the pure efficiency-ranked curve. Key for magnetization curves: MC = Monte Carlo, B+1 = best+1, ERP = pure efficiency-ranked, ERM = mixed efficiency-ranked, EX = exhausive search. }
\label{fig:lung_A549_p_eq_1_unc}
\end{figure}

Fig. \ref{fig:lung_A549_p_eq_1_unc} shows the results for the unconstrained $p=1$ model of the IMR-90/A549 lung cell network. (All simulations were performed using MATLAB on a desktop computer. Running the simulations to make all curves shown below required approximately 12 hours.) The unconstrained $p=1$ system has the largest search space, so the Monte Carlo strategy performs poorly. The best+1 strategy is the most effective strategy for controlling this network. The seed set of nodes used here was simply the size 1 bottleneck with the largest impact. Note that best+1 works better than effeciency-ranked. This is because best+1 includes the synergistic effects of fixing multiple nodes, while efficiency-ranked assumes that there is no overlap between the set of nodes downstream from multiple bottlenecks. Importantly, however, the efficiency-ranked method works nearly as well as best+1 and much better than Monte Carlo, both of which are more computationally expensive than the efficiency-ranked strategy.

\begin{figure}[!ht]
\begin{center}
\includegraphics[width=0.5\textwidth]{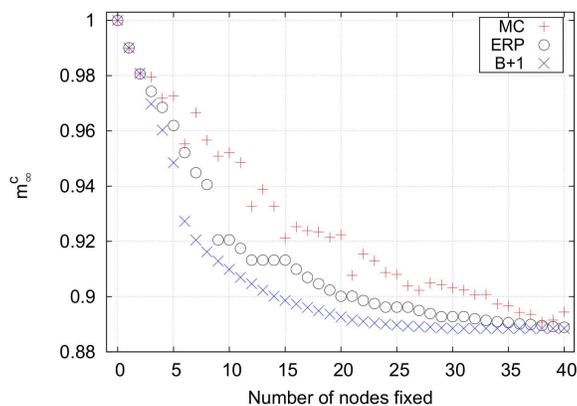}
\end{center}
\caption{{\bf Final cancer magnetizations for an unconstrained search on the lung cell network using $p=2$.} As in the $p=1$ case, the efficiency-ranked strategy outperforms the expensive Monte Carlo search. The plateaus in the efficiency-ranked strategy when fixing 9-10, 12-15, 20-21, etc. nodes are a result of targeting bottlenecks that are already indirectly controlled.}
\label{fig:lung_A549_p_eq_2_unc}
\end{figure}
\begin{figure}[!ht]
\begin{center}
\includegraphics[width=0.5\textwidth]{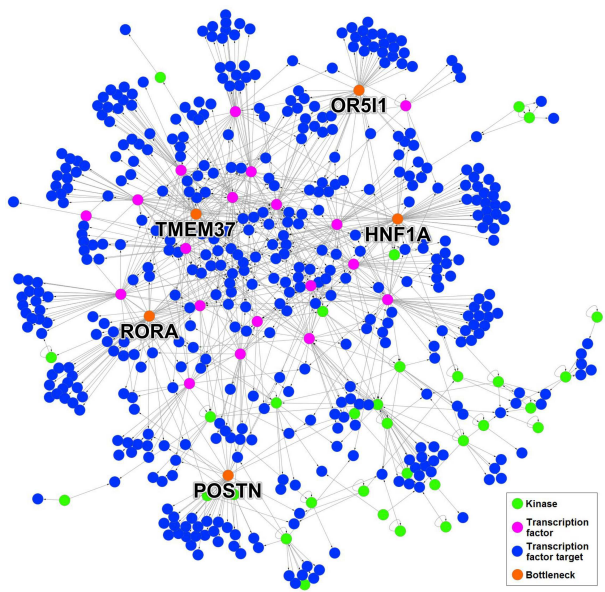}
\end{center}
\caption{{\bf Largest weakly connected differential subnetwork for IMR-90/A549 and $p=2$.} Out of the 506 pictured nodes, 450 are sinks and therefore have an impact equal to one. The top five bottlenecks are labeled with their gene names and colored orange.}
\label{fig:LungDiffSubnet_p_eq_2_IMR90_A549}
\end{figure}

Fig. \ref{fig:lung_A549_p_eq_2_unc} shows the results for the unconstrained $p=2$ model of the IMR-90/A549 lung cell network. The search space for $p=2$ is much smaller than that for $p=1$. The largest weakly connected differential subnetwork contains only 506 nodes (see Table \ref{table:model_2}) , and the remaining differential nodes are islets or are in subnetworks composed of two nodes and are therefore unnecessary to consider. Of these 506 nodes, 450 are sinks. Fig. \ref{fig:LungDiffSubnet_p_eq_2_IMR90_A549} shows the largest weakly connected component of the differential subnetwork, and the top five bottlenecks in the unconstrained case are shown in red. If limiting the search to differential kinases with $\xi^c_i=+1$ and ignoring all sinks, $p=2$ has 19 possible targets. There is only one cycle cluster in the largest differential subnetwork, containing 6 nodes. Like the $p=1$ case, the optimal efficiency occurs when targeting the first node, which is the highest impact size 1 bottleneck. Because the mixed efficiency-ranked strategy gives the same results as the pure efficiency-ranked strategy, only the pure strategy was examined. The Monte Carlo strategy fares better in the unconstrained $p=2$ case because the search space is smaller. Additionally, the efficiency-ranked strategy does worse against the best+1 strategy for $p=2$ than it did for $p=1$. This is because the effective edge deletion decreases the average indegree of the network and makes nodes easier to control indirectly. When many upstream bottlenecks are controlled, some of the downstream bottlenecks in the efficiency-ranked list can be indirectly controlled. Thus, controlling these nodes directly results in no change in the magnetization. This gives the plateaus shown for fixing nodes 9-10 and 12-15, for example.

\begin{figure}[!ht]
\begin{center}
\includegraphics[width=0.5\textwidth]{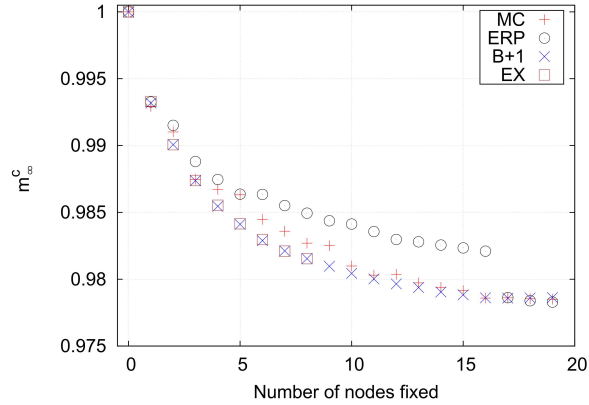}
\end{center}
\caption{{\bf Final cancer magnetizations for a constrained search on the lung cell network using $p=2$.} This is the only case in which a limited exhaustive search is possible. Interestingly, the exhaustive search locates the same nodes as the best+1 strategy for fixing up to eight nodes. The efficiency-ranked strategy performs poorly compared to the Monte Carlo strategy because the search space is small and a large portion of the available space is sampled by the Monte Carlo search.}
\label{fig:lung_A549_p_eq_2_con}
\end{figure}

The only case in which an exhaustive search is possible is for $p=2$ with constraints, which is shown in Fig. \ref{fig:lung_A549_p_eq_2_con}. Note that the polynomial-time best+1 strategy identifies the same set of nodes as the exponential-time exhaustive search. This is not surprising, however, since the constraints limit the available search space. This means that the Monte Carlo also does well. The efficiency-ranked method performs worst. The efficiency-ranked strategy is designed to be a heuristic strategy that scales gently, however, and is not expected to work well in such a small space when compared with more computationally expensive methods.




\subsection*{B Cell Network}
The B cell network was derived from the B cell interactome of Ref.~\cite{lefebvre2010human}. The reconstruction method used in Ref.~\cite{lefebvre2010human} removes edges from an initially complete network depending on pairwise gene expression correlation. 
Additionally, the original B cell network contains many protein-protein interactions (PPIs) as well as transcription factor-gene interactions (TFGIs). TFGIs have definite directionality: a transcription factor encoded by one gene affects the expression level of its target gene(s). PPIs, however, do not have obvious directionality. We first filtered these PPIs by checking if the genes encoding these proteins interacted according to the PhosphoPOINT/TRANSFAC network of the previous section, and if so, kept the edge as directed. If the remaining PPIs are ignored, the results for the B cell are similar to those of the lung cell network. We found more interesting results when keeping the remaining PPIs as undirected, as is discussed below.

Because of the network construction algorithm and the inclusion of many undirected edges, the B cell network is more dense ($\sim$0.290\% complete, see Table~\ref{table:general}) than the lung cell network. This higher density leads to many more cycles than the lung cell network, and many of these cycles overlap to form one very large cycle cluster containing $\sim$66\% of nodes in the full network. All gene expression data used for B cell attractors was taken from Ref.~\cite{compagno2009mutations}. We analyzed two types of normal B cells (na\"{\i}ve and memory) and three types of B cell cancers (diffuse large B-cell lymphoma (DLBCL), follicular lymphoma, and EBV-immortalized lymphoblastoma), giving six combinations in total. We present results for only the na\"{\i}ve/DLBCL combination below, but Tables~\ref{table:model_2} and \ref{table:best_singles} list the properties of all normal/cancer combinations. Again, all gene expression measurements for a given cell type were averaged together to produce a single attractor. The full B cell network is composed of 4364 nodes. For $p=1$, there is one cycle cluster $C$ composed of 2886 nodes. This cycle cluster has $1\leq \ncrit(C) \leq 1460$, $I(C)=4353$, and $3.0\leq \ecrit(C) \leq4353$. Finding $Z(C)$ was deemed too difficult. 

\begin{figure}[!ht]
\begin{center}
\includegraphics[width=0.5\textwidth]{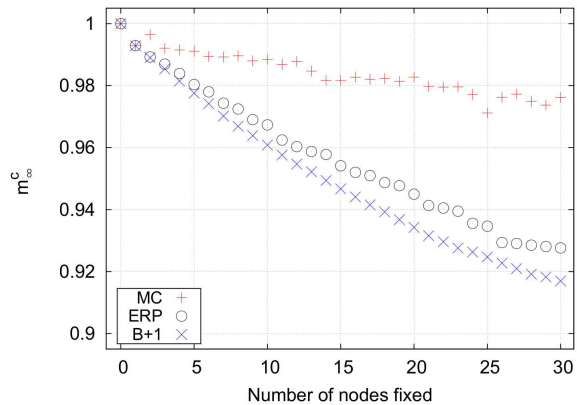}
\end{center}
\caption{{\bf Final cancer magnetizations for an unconstrained search on the B cell network using $p=1$.} The Monte Carlo strategy is ineffective for fixing any number of nodes. The efficiency-ranked and best+1 curves slowly separate because synergistic effects accumulate faster for best+1.}
\label{fig:b_p_eq_1_unc}
\end{figure}

Fig.\ref{fig:b_p_eq_1_unc} shows the results for the unconstrained $p=1$ case. Again, the pure efficiency-ranked strategy gave the same results as the mixed efficiency-ranked strategy, so only the pure strategy was analyzed. As shown in Fig.~\ref{fig:b_p_eq_1_unc}, the Monte Carlo strategy is out-performed by both the efficiency-ranked and best+1 strategies. The synergistic effects of fixing multiple bottlenecks slowly becomes apparent as the best+1 and efficiency-ranked curves separate.

\begin{figure}[!ht]
\begin{center}
\includegraphics[width=0.5\textwidth]{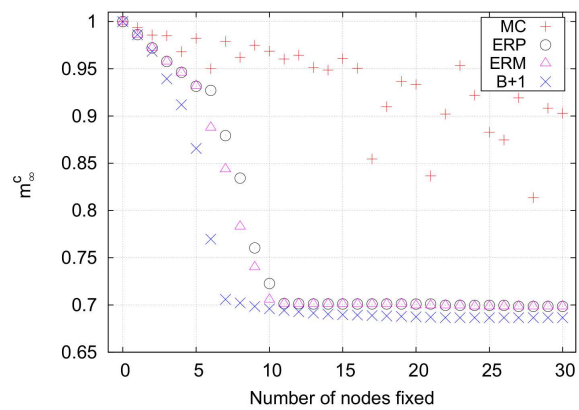}
\end{center}
\caption{{\bf Final cancer magnetizations for an unconstrained search on the B cell network using $p=2$.} The rather sudden drop in the magnetization between controlling 5 and 10 nodes in the efficiency-ranked strategies comes from flipping a significant portion of a cycle cluster. This is the only network examined in which the mixed efficiency-ranked strategy produces results different from the pure efficiency-ranked strategy.}
\label{fig:b_p_eq_2_unc}
\end{figure}

Fig. \ref{fig:b_p_eq_2_unc} shows the results for the unconstrained $p=2$ case. The largest weakly connected subnetwork contains one cycle cluster with 351 nodes, with $1\leq\ncrit\leq208$. Although finding a set of critical nodes is difficult, the optimal efficiency for this cycle cluster is 62.2 for fixing 10 bottlenecks in the cycle cluster. This makes targeting the cycle cluster worthwhile. The efficiency of this set of 10 nodes is larger than the efficiencies of the first 10 nodes from the pure efficiency-ranked strategy, so the $\mcinf$ from the mixed strategy drops earlier than the pure strategy. Both strategies quickly identify a small set of nodes capable of controlling a significant portion of the differential network, however, and the same result is obtained for fixing more than 10 nodes. The best+1 strategy finds a smaller set of nodes that controls a similar fraction of the cycle cluster, and fixing more than 7 nodes results in only incremental decreases in $\mcinf$. The Monte Carlo strategy performs poorly, never finding a set of nodes adequate to control a significant fraction of the nodes in the cycle cluster.
%


\section*{Conclusion}
\label{sec:conclusions}
Signaling models for  large and complex biological networks are becoming important tools for designing new therapeutic methods for complex diseases such as cancer. Even if our knowledge of biological networks is incomplete, rapid progress is currently being made using reconstruction methods that use large amounts of publicly available omic data~\cite{de2010advantages,hartemink2005reverse}. The Hopfield model we use in our approach allows mapping of gene expression patterns of normal and cancer cells into stored attractor states of the signaling dynamics in directed networks. The role of each node in disrupting the network signaling can therefore be explicitly analyzed to identify isolated genes or sets of strongly connected genes that are selective in their action. We have introduced the concept of \emph{size $k$ bottlnecks} to identify such genes. This concept led to the formulation of several heuristic strategies, such as the \emph{efficiency-ranked} and \emph{best+1} strategy to find nodes that reduce the overlap of the cell network with a cancer attractor. Using this approach, we have located small sets of nodes in lung and B cancer cells which, when forced away from their initial states with local magnetic fields (representing targeted drugs), disrupt the signaling of the cancer cells while leaving normal cells in their original state. For networks with few targetable nodes, exhaustive searches or Monte Carlo searches can locate effective sets of nodes. For larger networks, however, these strategies become too cumbersome and our  heuristic strategies represent a feasible alternative. For tree-like networks, the pure efficiency-ranked strategy works well, whereas the mixed efficiency-ranked strategy could be a better choice for networks with high-impact cycle clusters. 

We make two important assumptions in applying this analysis to real biological systems. First, we assume that genes are either fully off or fully on, with no intermediate state. Modelling the state of a neuron as ``all-or-none'' has long been accepted as a reasonable assumption~\cite{McCullochPittsNeurons1943}, which provided the spin glass framework for the Hopfield model. While similar switch-like behavior in gene regulatory networks has been proposed as an explanation of, for example, segmentation in \emph{Drosophila} embryos~\cite{Krotov10022014}, assigning a Boolean value to gene expression may be overly simplistic in many cases. A model which uses spins with more than two projections could prove to be more realistic and predictive. Second, we assume that all nodes update their status with a single timescale and with a single interaction strength. If the signaling timescale $\tau_{ij}$ for each edge in the biological network is known, simulations could be conducted in which a signal traveling along an edge $(j,i)$ reaches its target after $\tau_{ij}$ time steps. This would amount to a synchronous update schedule with a ``queue'' of signals moving between nodes. Likewise, our model gives equal weight to all edges (aside from edges that are effectively deleted in the $p=2$ case), whereas real gene regulatory networks exhibit a spectrum of interaction strengths. This can easily be integrated with our model by using a weighted, directed adjacency matrix. However, doing this would surely require a change in control strategy.

Despite these issues, our model shows promise. Some of the genes identified in Table \ref{table:best_singles} are consistent with current clinical and cancer biology knowledge. For instance, in the lung cancer list we found a well known tumor suppressor gene (TP53)~\cite{baker1989chromosome} that is frequently mutated in many cancer types including lung cancer~\cite{takahashi1989p53}. Mutations in PBX1 have recently been detected in non-small-cell lung cancer and this gene is now being considered as a target for therapy and prognosis~\cite{mo2013detection}. MAP3K3 and MAP3K14 are in the MAPK/ERK pathway which is a target of many novel therapeutic agents~\cite{montagut2009targeting}, and SRC is a well known oncogene and a candidate target in lung cancer~\cite{rothschild2010src}. BCL6 (B-cell lymphoma 6) is the most common oncogene in DLBCL, and it is known that its expression can predict prognosis and response to drug therapy~\cite{hans2004confirmation,rosenwald2002use,winter2006prognostic}.  BCL6 is also frequently mutated in follicular lymphoma ~\cite{diaz2008frequency,akasaka2003bcl6}.  Our analysis identified BCL6 as an important drug target for both DLBCL and follicular lymphoma using either naive or  
memory B-cells as a control for both $p=1$ and $p=2$. RBL2 disregulation has been recently associated with many types of lymphoma~\cite{wang2008protein,de2007gene,piccaluga2011gene}. FOXM1 is a potential therapeutic target in mature B cell tumors~\cite{tompkins2013identification} and  ATF2 has been recently found to be highly disregulated in lymphoma~\cite{valdez2014synergistic,walczynski2013sensitisation}. Besides BCL6 discussed above, the N/D list for DLBCL contains genes (MEF2A~\cite{bai2008overexpression}, NCOA1~\cite{fabris2013chromosome,zhang2013steroid}, TGIF1~\cite{hamid2009transforming,liborio2013tgif1,bengoechea2010tumor}, NFATC3~\cite{glud2005tumor}) that are all known to have a functional role in cancer, even if they have not been  associated to the specific  B-cell cancer types we have considered. Our predictions are for the immortalized cell lines we have selected, some of which are commonly used for in-vitro testing in many laboratories.  RNAi and targeted drugs could then be used  in these cell lines against the top scoring genes in Table \ref{table:best_singles} to test the disruption of survival or proliferative capacity. If experimentally validated, our analysis based on attractor states and bottlenecks could be applied to patient-derived cancer cells by integrating in the model patient gene expression data to identify patient-specific targets.

The above unconstrained searches assume that there exists some set of ``miracle drugs'' which can turn any gene ``on'' and ``off'' at will. This limitation can be patially taken into account by using constrained searches that limit the nodes that can be addressed. However, even the constrained search results are unrealistic, since most drugs directly target more than one gene. Inhibitors, for example, could target differential nodes with $\xi^c_i=-1$ and $\xi^n_i=+1$, which would damage only normal cells. Additionally, drugs would not be restricted to target only differential nodes, and certain combinations could be toxic to both normal and cancer cells. Few cancer treatments involve the use of a single drug, and the synergistic effects of combining multiple drugs adds yet another level of complication to finding an effective treatment~\cite{feala2012statistical}. On the other hand, the intrinsic nonlinearity of a cellular signaling network, with its inherent structure of attractor states, enhances control~\cite{cornelius2013realistic} so that a properly selected set of druggable targets might be sufficient for robust control.


\section*{Acknowledgments}


We thank Andrew Hodges and Jacob Feala for help with biological datasets. Correspondence and requests for materials should be addressed to carlo@pa.msu.edu or szedlak1@msu.edu.


\bibliography{bibcarlo2}

\begin{thebibliography}{10}

\bibitem{agoston2005multiple}
V.~{\'A}goston, P.~Csermely, and S.~Pongor.
\newblock Multiple weak hits confuse complex systems: a transcriptional
  regulatory network as an example.
\newblock {\em Phys. Rev. E}, 71(5):051909, 2005.

\bibitem{akasaka2003bcl6}
T.~Akasaka, I.S. Lossos, and R.~Levy.
\newblock Bcl6 gene translocation in follicular lymphoma: a harbinger of
  eventual transformation to diffuse aggressive lymphoma.
\newblock {\em Blood}, 102(4):1443--1448, 2003.

\bibitem{akutsu2007control}
T.~Akutsu, M.~Hayashida, W.K. Ching, and M.K. Ng.
\newblock Control of boolean networks: hardness results and algorithms for tree
  structured networks.
\newblock {\em J. Theor. Biol.}, 244(4):670--679, 2007.

\bibitem{aldana2003boolean}
M.~Aldana, S.~Coppersmith, and L.P. Kadanoff.
\newblock Boolean dynamics with random couplings.
\newblock In {\em Perspectives and Problems in Nonlinear Sciences}, pages
  23--89. Springer, 2003.

\bibitem{PhysRevE.87.022814}
S.~Amari, H.~Ando, T.~Toyoizumi, and N.~Masuda.
\newblock State concentration exponent as a measure of quickness in
  kauffman-type networks.
\newblock {\em Phys. Rev. E}, 87:022814, Feb 2013.

\bibitem{amit1985spin}
D.J. Amit, H.~Gutfreund, and H.~Sompolinsky.
\newblock Spin-glass models of neural networks.
\newblock {\em Phys. Rev. A}, 32(2):1007, 1985.

\bibitem{10.1371/journal.pone.0014413}
Ron~C. Anafi and Jason H.~T. Bates.
\newblock Balancing robustness against the dangers of multiple attractors in a
  hopfield-type model of biological attractors.
\newblock {\em PLoS ONE}, 5(12):e14413, 12 2010.

\bibitem{ao2008cancer}
P.~Ao, D.~Galas, L.~Hood, and X.~Zhu.
\newblock Cancer as robust intrinsic state of endogenous molecular-cellular
  network shaped by evolution.
\newblock {\em Med. Hypotheses}, 70(3):678--684, 2008.

\bibitem{bai2008overexpression}
X.~Bai, L.~Wu, T.~Liang, Z.~Liu, J.~Li, D.~Li, H.~Xie, S.~Yin, J.~Yu, Q.~Lin,
  et~al.
\newblock Overexpression of myocyte enhancer factor 2 and histone
  hyperacetylation in hepatocellular carcinoma.
\newblock {\em J. Canc. Res. Clinic. Oncol.}, 134(1):83--91, 2008.

\bibitem{baker1989chromosome}
S.J. Baker, E.R. Fearon, J.~M. Nigro, A.C. Preisinger, J.M. Jessup, D.H.
  Ledbetter, D.F. Barker, Y.~Nakamura, R.~White, B.~Vogelstein, et~al.
\newblock Chromosome 17 deletions and p53 gene mutations in colorectal
  carcinomas.
\newblock {\em Science}, 244(4901):217--221, 1989.

\bibitem{bengoechea2010tumor}
M.T. Bengoechea-Alonso and J.~Ericsson.
\newblock Tumor suppressor fbxw7 regulates tgf$\beta$ signaling by targeting
  tgif1 for degradation.
\newblock {\em Oncogene}, 29(38):5322--5328, 2010.

\bibitem{bhardwaj2010analysis}
N.~Bhardwaj, M.B. Carson, A.~Abyzov, K.-K. Yan, H.~Lu, and M.B. Gerstein.
\newblock Analysis of combinatorial regulation: scaling of partnerships between
  regulators with the number of governed targets.
\newblock {\em PLoS Comp. Biol.}, 6(5):e1000755, 2010.

\bibitem{bullinger2004use}
L.~Bullinger, K.~D{\"o}hner, E.~Bair, S.~Fr{\"o}hling, R.F. Schlenk,
  R.~Tibshirani, H.~D{\"o}hner, and J.R. Pollack.
\newblock Use of gene-expression profiling to identify prognostic subclasses in
  adult acute myeloid leukemia.
\newblock {\em New Engl. J. Med.}, 350(16):1605--1616, 2004.

\bibitem{calzolari2007selective}
D.~Calzolari, G.~Paternostro, P.L. Harrington~Jr., C.~Piermarocchi, and P.M.
  Duxbury.
\newblock Selective control of the apoptosis signaling network in heterogeneous
  cell populations.
\newblock {\em PLoS ONE}, 2(6):e547, 2007.

\bibitem{choudhary2006intervention}
A.~Choudhary, A.~Datta, M.L. Bittner, and E.R. Dougherty.
\newblock Intervention in a family of boolean networks.
\newblock {\em Bioinformatics}, 22(2):226--232, 2006.

\bibitem{cohen2002protein}
P.~Cohen.
\newblock Protein kinases - the major drug targets of the twenty-first century?
\newblock {\em Nature Rev. Drug Discov.}, 1(4):309--315, 2002.

\bibitem{compagno2009mutations}
Mara Compagno, Wei~Keat Lim, Adina Grunn, Subhadra~V Nandula, Manisha
  Brahmachary, Qiong Shen, Francesco Bertoni, Maurilio Ponzoni, Marta
  Scandurra, Andrea Califano, et~al.
\newblock Mutations of multiple genes cause deregulation of nf-$\kappa$b in
  diffuse large b-cell lymphoma.
\newblock {\em Nature}, 459(7247):717--721, 2009.

\bibitem{cornelius2013realistic}
S.P. Cornelius, W.L. Kath, and A.E Motter.
\newblock Realistic control of network dynamics.
\newblock {\em Nature Commun.}, 4:1--9, 2013.

\bibitem{creixell2012navigating}
P.~Creixell, E.~M Schoof, J.T. Erler, and R.~Linding.
\newblock Navigating cancer network attractors for tumor-specific therapy.
\newblock {\em Nature Biotechnol.}, 30(9):842--848, 2012.

\bibitem{csermely2005efficiency}
P{\'e}ter Csermely, Vilmos {\'A}goston, and Sandor Pongor.
\newblock The efficiency of multi-target drugs: the network approach might help
  drug design.
\newblock {\em Trends in Pharmacological Sciences}, 26(4):178--182, 2005.

\bibitem{de2007gene}
G.~De~Falco, E.~Leucci, D.~Lenze, P.P. Piccaluga, P.P. Claudio, A.~Onnis,
  G.~Cerino, J.~Nyagol, W.~Mwanda, C.~Bellan, et~al.
\newblock Gene-expression analysis identifies novel rbl2/p130 target genes in
  endemic burkitt lymphoma cell lines and primary tumors.
\newblock {\em Blood}, 110(4):1301--1307, 2007.

\bibitem{de2010advantages}
R.~De~Smet and K.~Marchal.
\newblock Advantages and limitations of current network inference methods.
\newblock {\em Nature Rev. Microbiol.}, 8(10):717--729, 2010.

\bibitem{derrida1987exactly}
B.~Derrida, E.~Gardner, and A.~Zippelius.
\newblock An exactly solvable asymmetric neural network model.
\newblock {\em Europhys. Lett.)}, 4(2):167, 1987.

\bibitem{diaz2008frequency}
A.~Diaz-Alderete, A.~Doval, F.~Camacho, L.~Verde, P.~Sabin, R.~Arranz-Saez,
  C.~Bellas, C.~Corbacho, J.~Gil, M.~Perez-Martin, et~al.
\newblock Frequency of bcl2 and bcl6 translocations in follicular lymphoma:
  relation with histological and clinical features.
\newblock {\em Leukemia Lymphoma}, 49(1):95--101, 2008.

\bibitem{eppert2011stem}
K.~Eppert, K.~Takenaka, E.R. Lechman, L.~Waldron, B.~Nilsson, P.~van Galen,
  K.H. Metzeler, A.~Poeppl, V.~Ling, J.~Beyene, et~al.
\newblock Stem cell gene expression programs influence clinical outcome in
  human leukemia.
\newblock {\em Nature Med.}, 17(9):1086--1093, 2011.

\bibitem{fabris2013chromosome}
S.~Fabris, L.~Mosca, G.~Cutrona, M.~Lionetti, L.~Agnelli, G.~Ciceri,
  M.~Barbieri, F.~Maura, S.~Matis, M.~Colombo, et~al.
\newblock Chromosome 2p gain in monoclonal b-cell lymphocytosis and in early
  stage chronic lymphocytic leukemia.
\newblock {\em Am. J. Hemat.}, 88(1):24--31, 2013.

\bibitem{Fagiolo2007clustering}
G.~Fagiolo.
\newblock Clustering in complex directed networks.
\newblock {\em Phys. Rev. E}, 76:026107, Aug 2007.

\bibitem{feala2012statistical}
J.D. Feala, J.~Cortes, P.M. Duxbury, A.D. McCulloch, C.~Piermarocchi, and
  G.~Paternostro.
\newblock Statistical properties and robustness of biological controller-target
  networks.
\newblock {\em PLoS ONE}, 7(1):e29374, 2012.

\bibitem{feala2010systems}
J.D. Feala, J.~Cortes, P.M. Duxbury, C.~Piermarocchi, A.D. McCulloch, and
  G.~Paternostro.
\newblock Systems approaches and algorithms for discovery of combinatorial
  therapies.
\newblock {\em Wiley Interdisciplinary Reviews: Systems Biology and Medicine},
  2(2):181--193, 2010.

\bibitem{girvan2002community}
M.~Girvan and M.~E.~J. Newman.
\newblock Community structure in social and biological networks.
\newblock {\em Proc. Nat. Acad. Sci. USA}, 99(12):7821--7826, 2002.

\bibitem{glud2005tumor}
S.~Z. Glud, A.~B. S{\"o}rensen, M.~Andrulis, B.~Wang, E.~Kondo, R.~Jessen,
  L.~Krenacs, E.~Stelkovics, M.~Wabl, E.~Serfling, et~al.
\newblock A tumor-suppressor function for nfatc3 in t-cell lymphomagenesis by
  murine leukemia virus.
\newblock {\em Blood}, 106(10):3546--3552, 2005.

\bibitem{hamid2009transforming}
R.~Hamid and S.J. Brandt.
\newblock Transforming growth-interacting factor tgif regulates proliferation
  and differentiation of human myeloid leukemia cells.
\newblock {\em Mol. Oncol.}, 3(5):451--463, 2009.

\bibitem{hans2004confirmation}
C.P. Hans, D.D. Weisenburger, T.C. Greiner, R.D. Gascoyne, J.~Delabie, G.~Ott,
  H.K. M{\"u}ller-Hermelink, E.~Campo, R.M. Braziel, E.~S. Jaffe, et~al.
\newblock Confirmation of the molecular classification of diffuse large b-cell
  lymphoma by immunohistochemistry using a tissue microarray.
\newblock {\em Blood}, 103(1):275--282, 2004.

\bibitem{hartemink2005reverse}
A.J. Hartemink.
\newblock Reverse engineering gene regulatory networks.
\newblock {\em Nature Biotechnol.}, 23(5):554--555, 2005.

\bibitem{hopfield1982neural}
J.J. Hopfield.
\newblock Neural networks and physical systems with emergent collective
  computational abilities.
\newblock {\em Proc. Nat. Acad. Sci. USA}, 79(8):2554--2558, 1982.

\bibitem{PhysRevLett.94.128701}
S.~Huang, G.~Eichler, Y.~Bar-Yam, and D.E. Ingber.
\newblock Cell fates as high-dimensional attractor states of a complex gene
  regulatory network.
\newblock {\em Phys. Rev. Lett.}, 94:128701, Apr 2005.

\bibitem{hussain2009tobacco}
Mustafa Hussain, Mahadev Rao, Ashley~E Humphries, Julie~A Hong, Fang Liu,
  Maocheng Yang, Diana Caragacianu, and David~S Schrump.
\newblock Tobacco smoke induces polycomb-mediated repression of dickkopf-1 in
  lung cancer cells.
\newblock {\em Cancer research}, 69(8):3570--3578, 2009.

\bibitem{Kauffman1969437}
S.A. Kauffman.
\newblock Metabolic stability and epigenesis in randomly constructed genetic
  nets.
\newblock {\em J. Theor. Biol.}, 22(3):437 -- 467, 1969.

\bibitem{Krotov10022014}
Dmitry Krotov, Julien~O. Dubuis, Thomas Gregor, and William Bialek.
\newblock Morphogenesis at criticality.
\newblock {\em Proceedings of the National Academy of Sciences}, 2014.

\bibitem{0305-4470-21-11-009}
K.E. K\"urten.
\newblock Correspondence between neural threshold networks and kauffman boolean
  cellular automata.
\newblock {\em J. Phys. A}, 21(11):L615, 1988.

\bibitem{Kürten1988157}
K.E. K\"urten.
\newblock Critical phenomena in model neural networks.
\newblock {\em Phys. Lett. A}, 129(3):157 -- 160, 1988.

\bibitem{2012arXiv1211.3133L}
A.~H. {Lang}, H.~{Li}, J.~J. {Collins}, and P.~{Mehta}.
\newblock {Epigenetic landscapes explain partially reprogrammed cells and
  identify key reprogramming genes}.
\newblock {\em ArXiv e-prints}, page arXiv:1211.3133v3, November 2012.

\bibitem{lefebvre2010human}
C.~Lefebvre, P.~Rajbhandari, M.J. Alvarez, P.~Bandaru, W.K. Lim, M.~Sato,
  K.~Wang, P.~Sumazin, M.~Kustagi, B.C Bisikirska, et~al.
\newblock A human b-cell interactome identifies myb and foxm1 as master
  regulators of proliferation in germinal centers.
\newblock {\em Mol. Syst. Biol.}, 6(1), 2010.

\bibitem{liborio2013tgif1}
T.N. Lib{\'o}rio, E.~N. Ferreira, F.~C. Aquino~Xavier, D.~M. Carraro, L.~P.
  Kowalski, F.~A. Soares, and F.D. Nunes.
\newblock Tgif1 splicing variant 8 is overexpressed in oral squamous cell
  carcinoma and is related to pathologic and clinical behavior.
\newblock {\em Oral Surg. Oral Med.}, 116(5):614--625, 2013.

\bibitem{liu2011controllability}
Yang-Yu Liu, Jean-Jacques Slotine, and Albert-L{\'a}szl{\'o} Barab{\'a}si.
\newblock Controllability of complex networks.
\newblock {\em Nature}, 473(7346):167--173, 2011.

\bibitem{gerstein2004topologicalchanges}
N.M. Luscombe, M.M. Babu, H.~Yu, M.~Snyder, S.A. Teichmann, and M.~Gerstein.
\newblock Genomic analysis of regulatory network dynamics reveals large
  topological changes.
\newblock {\em Nature}, 431(7006):308--312, 2004.

\bibitem{Matys01012003}
V.~Matys, E.~Fricke, R.~Geffers, E.~G\"ossling, M.~Haubrock, R.~Hehl,
  K.~Hornischer, D.~Karas, A.~E. Kel, O.~V. Kel-Margoulis, et~al.
\newblock Transfac: transcriptional regulation, from patterns to profiles.
\newblock {\em Nucleic Acids Res.}, 31(1):374--378, 2003.

\bibitem{McCullochPittsNeurons1943}
WarrenS. McCulloch and Walter Pitts.
\newblock A logical calculus of the ideas immanent in nervous activity.
\newblock {\em The bulletin of mathematical biophysics}, 5(4):115--133, 1943.

\bibitem{mo2013detection}
M.-L. Mo, Z.~Chen, H.-M. Zhou, H.~Li, T.~Hirata, D.M. Jablons, and B.~He.
\newblock Detection of e2a-pbx1 fusion transcripts in human non-small-cell lung
  cancer.
\newblock {\em J. Exp. Clin. Canc. Res.}, 32(1):29, 2013.

\bibitem{montagut2009targeting}
C.~Montagut and J.~Settleman.
\newblock Targeting the raf--mek--erk pathway in cancer therapy.
\newblock {\em Canc. Lett.}, 283(2):125--134, 2009.

\bibitem{muggerud2009data}
Aslaug~A Muggerud, Henrik Edgren, Maija Wolf, Kristine Kleivi, Emelyne Dejeux,
  J{\"o}rg Tost, Therese S{\o}rlie, and Olli Kallioniemi.
\newblock Data integration from two microarray platforms identifies bi-allelic
  genetic inactivation of ric8a in a breast cancer cell line.
\newblock {\em BMC medical genomics}, 2(1):26, 2009.

\bibitem{munoz2011quantitative}
Javier Munoz, Teck~Y Low, Yee~J Kok, Angela Chin, Christian~K Frese, Vanessa
  Ding, Andre Choo, and Albert~JR Heck.
\newblock The quantitative proteomes of human-induced pluripotent stem cells
  and embryonic stem cells.
\newblock {\em Molecular systems biology}, 7(1), 2011.

\bibitem{muzikar2009repression}
Katy~A Muzikar, Nicholas~G Nickols, and Peter~B Dervan.
\newblock Repression of dna-binding dependent glucocorticoid receptor-mediated
  gene expression.
\newblock {\em Proceedings of the National Academy of Sciences},
  106(39):16598--16603, 2009.

\bibitem{piccaluga2011gene}
P.P. Piccaluga, G.~De~Falco, M.~Kustagi, A.~Gazzola, C.~Agostinelli,
  C.~Tripodo, E.~Leucci, A.~Onnis, A.~Astolfi, M.~R. Sapienza, et~al.
\newblock Gene expression analysis uncovers similarity and differences among
  burkitt lymphoma subtypes.
\newblock {\em Blood}, 117(13):3596--3608, 2011.

\bibitem{plummer1986matching}
M.~D. Plummer and L.~Lov{\'a}sz.
\newblock {\em Matching theory}.
\newblock Elsevier, 1986.

\bibitem{ravasz2002hierarchical}
E.~Ravasz, A.L. Somera, D.A. Mongru, Z.N. Oltvai, and A.-L. Barab{\'a}si.
\newblock Hierarchical organization of modularity in metabolic networks.
\newblock {\em Science}, 297(5586):1551--1555, 2002.

\bibitem{rohlf2009self}
T.~Rohlf and S.~Bornholdt.
\newblock Self-organized criticality and adaptation in discrete dynamical
  networks.
\newblock In {\em Adaptive Networks}, pages 73--106. Springer, 2009.

\bibitem{rosenwald2002use}
A.~Rosenwald, G.~Wright, W.C. Chan, J.~M. Connors, E.~Campo, R.I. Fisher, R.D.
  Gascoyne, H.K. Muller-Hermelink, E.B. Smeland, J.~M. Giltnane, et~al.
\newblock The use of molecular profiling to predict survival after chemotherapy
  for diffuse large-b-cell lymphoma.
\newblock {\em New Engl. J. Med.}, 346(25):1937--1947, 2002.

\bibitem{rothschild2010src}
S.I. Rothschild, O.~Gautschi, E.B. Haura, and F.M. Johnson.
\newblock Src inhibitors in lung cancer: current status and future directions.
\newblock {\em Clin. Lung Canc.}, 11(4):238--242, 2010.

\bibitem{sartor2010conceptgen}
Maureen~A Sartor, Vasudeva Mahavisno, Venkateshwar~G Keshamouni, James
  Cavalcoli, Zachary Wright, Alla Karnovsky, Rork Kuick, HV~Jagadish, Barbara
  Mirel, Terry Weymouth, et~al.
\newblock Conceptgen: a gene set enrichment and gene set relation mapping tool.
\newblock {\em Bioinformatics}, 26(4):456--463, 2010.

\bibitem{shen2002network}
S.S. Shen-Orr, R.~Milo, S.~Mangan, and U.~Alon.
\newblock Network motifs in the transcriptional regulation network of
  escherichia coli.
\newblock {\em Nature Genet.}, 31(1):64--68, 2002.

\bibitem{sontag1998mathematical}
E.D. Sontag.
\newblock {\em Mathematical control theory: deterministic finite dimensional
  systems}, volume~6.
\newblock Springer, 1998.

\bibitem{stinson2011trps1}
Susanna Stinson, Mark~R Lackner, Alex~T Adai, Nancy Yu, Hyo-Jin Kim, Carol
  O'Brien, Jill Spoerke, Suchit Jhunjhunwala, Zachary Boyd, Thomas Januario,
  et~al.
\newblock Trps1 targeting by mir-221/222 promotes the epithelial-to-mesenchymal
  transition in breast cancer.
\newblock {\em Science Signaling}, 4(177):ra41, 2011.

\bibitem{Huang2009869}
H.~Sui, I.~Ernberg, and S.~Kauffman.
\newblock Cancer attractors: A systems view of tumors from a gene network
  dynamics and developmental perspective.
\newblock {\em Sem. Cell Dev. Biol.}, 20(7):869 -- 876, 2009.

\bibitem{takahashi1989p53}
T.~Takahashi, M.M. Nau, I.~Chiba, M.J. Birrer, R.K. Rosenberg, M.~Vinocour,
  M.~Levitt, H.~Pass, A.F. Gazdar, and J.D. Minna.
\newblock p53: a frequent target for genetic abnormalities in lung cancer.
\newblock {\em Science}, 246(4929):491--494, 1989.

\bibitem{tompkins2013identification}
V.S. Tompkins, S.-S. Han, A.~Olivier, S.~Syrbu, T.~Bair, A.~Button, Laura
  Jacobus, Zebin Wang, Samuel Lifton, Pradip Raychaudhuri, et~al.
\newblock Identification of candidate b-lymphoma genes by cross-species gene
  expression profiling.
\newblock {\em PLoS ONE}, 8(10):e76889, 2013.

\bibitem{valdez2014synergistic}
B.C. Valdez, A.R. Zander, G.~Song, D.~Murray, Y.~Nieto, Y.~Li, R.E. Champlin,
  and B.S. Andersson.
\newblock Synergistic cytotoxicity of gemcitabine, clofarabine and edelfosine
  in lymphoma cell lines.
\newblock {\em Blood Canc. J.}, 4(1):e171, 2014.

\bibitem{wagner2007death}
Klaus~W Wagner, Elizabeth~A Punnoose, Thomas Januario, David~A Lawrence,
  Robert~M Pitti, Kate Lancaster, Dori Lee, Melissa von Goetz, Sharon~Fong Yee,
  Klara Totpal, et~al.
\newblock Death-receptor o-glycosylation controls tumor-cell sensitivity to the
  proapoptotic ligand apo2l/trail.
\newblock {\em Nature medicine}, 13(9):1070--1077, 2007.

\bibitem{walczynski2013sensitisation}
J.~Walczynski, S.~Lyons, N.~Jones, and W.~Breitwieser.
\newblock Sensitisation of c-myc-induced b-lymphoma cells to apoptosis by atf2.
\newblock {\em Oncogene}, 33:1027--1036, 2013.

\bibitem{wang2008protein}
L.~Wang, S.~Pal, and S.~Sif.
\newblock Protein arginine methyltransferase 5 suppresses the transcription of
  the rb family of tumor suppressors in leukemia and lymphoma cells.
\newblock {\em Mol. Cell. Biol.}, 28(20):6262--6277, 2008.

\bibitem{winter2006prognostic}
J.N. Winter, E.A. Weller, S.J. Horning, M.~Krajewska, D.~Variakojis, T.M.
  Habermann, R.I. Fisher, P.J. Kurtin, W.R. Macon, M.~Chhanabhai, et~al.
\newblock Prognostic significance of bcl-6 protein expression in dlbcl treated
  with chop or r-chop: a prospective correlative study.
\newblock {\em Blood}, 107(11):4207--4213, 2006.

\bibitem{Yang15082008}
C.-Y. Yang, C.-H. Chang, Y.-L. Yu, T.-C.~E. Lin, S.-A. Lee, C.-C. Yen, J.-M.
  Yang, J.-M. Lai, Y.-R. Hong, T.-L. Tseng, K.-M. Chao, and C.-Y.~F. Huang.
\newblock Phosphopoint: a comprehensive human kinase interactome and
  phospho-protein database.
\newblock {\em Bioinformatics}, 24(16):i14--i20, 2008.

\bibitem{zhang2013steroid}
Y.~Zhang, C.~Duan, C.~Bian, Y.~Xiong, and J.~Zhang.
\newblock Steroid receptor coactivator-1: A versatile regulator and promising
  therapeutic target for breast cancer.
\newblock {\em J. Steroid Biochem.}, 138:17, 2013.

\end{thebibliography}


\newpage

\section*{Tables}

\begin{table}[!ht]
\centering
  \begin{tabular}{ l | l }
	Symbol & Explanation \\ \hline

   $G$  &  Set of nodes and directed edges (network) \\
   $N$  &  Number of nodes \\
   $A_{ij}$  &  Adjacency matrix \\
   $V(G)$  &  Set of nodes in $G$ \\
   $E(G)$  &  Set of edges in $G$ \\
   $\text{deg}^{+/-}(i)$  &  Outdegree/indegree of node $i$ \\
   $\sigma_i$  &  Spin of node $i$, $=\pm1$ \\
   $\xi^a$  &  $a^{\text{th}}$ attractor \\
   $\xi^{n/c}$  &  Normal/cancer attractor \\
   $J_{ij}$  &  Coupling matrix \\
   $h_i$  &  Total field at node $i$ \\
   $h_i^{\text{ext}}$  &  External field applied to node $i$ \\
   $T$  &  Temperature \\
   $Q$  &  Set of source and effective source nodes \\
   $m^a(t)$  &  Magnetization along attractor $a$ at time $t$\\
   $m^a_{\infty}$  &  Steady-state magnetization along attractor $a$ \\
   $p$  &  Number of attractors in coupling matrix \\
   $S$  &  Set of similarity nodes \\
   $D$  &  Set of differential nodes \\
   $L(B)$  &  Control set of bottleneck $B$ \\
   $I(B)$  &  Impact of bottleneck $B$ \\
   $C$  &  Cycle cluster \\
   $B$  &  Size $k$ bottleneck, where $k=|B|$ \\
   $Z(B,G)$  &  Set of critical nodes for bottleneck $B$ in network $G$ \\
   $\ncrit(B,G)$  &  Critical number of nodes in bottleneck $B$ in network $G$ \\
   $R(C,G)$  &  Set of externally influenced nodes \\
   $W(C,G)$  &  Set of intruder connections \\
   $Z_{\text{red}}(C,G)$  &  Reduced set of critical nodes \\
   $\mu$  &  Minimum indegree of all nodes in a cycle cluster \\
   $\ecrit(B)$  &  Critical efficiency of bottleneck $B$ \\
   $\eopt(B)$  &  Optimal efficiency of bottleneck $B$
  \end{tabular}
\caption{{\bf Reference table for symbols.} This table lists all important symbols introduced in the article with a brief explanation of its purpose.
\label{table:symbols}}
\end{table}


\begin{table}[!ht]
\centering
  \begin{tabular}{ l | c | c }
	Properties & Lung & B cell \\ \hline
	Nodes & 9073 & 4364 \\
	Edges & 45635 & 55144 \\
	Sources & 129 & 8 \\
	Sinks & 8443 & 1418 \\
	Av. outdegree & 5.03 & 12.64 \\
	Max outdegree & 240 & 2372 \\
	Max indegree & 68 & 196 \\
	Self-loops & 238 & 0 \\
	Undirected edges & 350 & 23386 \\
	Diameter & 11 & 11 \\
	Max cycle cluster & 401 & 2886 \\
	Av. clustering coeff.~\cite{Fagiolo2007clustering} & 0.0544 & 0.2315
  \end{tabular}
\caption{{\bf General properties of the full networks.} The network used for the analysis of lung cancer is a generic one obtained combining the data sets in Refs.~\cite{Yang15082008} and \cite{Matys01012003}. The B cell network is a curated version of the B cell interactome obtained in Ref.~\cite{lefebvre2010human} using a network reconstruction method and gene expression data from B cells. 
\label{table:general}}
\end{table}




\begin{table*}[!ht]
\centering
\begin{tabular}{ l | r r | r r r r r r }
\multirow{2}{*}[-0.3cm]{Properties} 	& \multicolumn{2}{c}{Lung} & \multicolumn{6}{c}{B} \\ 
									& \multicolumn{1}{c}{I/A} & \multicolumn{1}{c|}{I/H} & \multicolumn{1}{c}{N/D} & \multicolumn{1}{c}{N/F} & \multicolumn{1}{c}{N/L} & \multicolumn{1}{c}{M/D} & \multicolumn{1}{c}{M/F} & \multicolumn{1}{c}{M/L} \\ \hline

Nodes 								 & 506 & 667 & 684 & 511 & 841 &  621 & 457 & 742 \\
Edges							 		 & 846 & 1227 & 2855 & 1717 & 3962 & 2525 & 1501 & 3401 \\
Sources and effective sources	 & 30 & 34 & 12 & 11 & 9 & 9 & 9 & 12 \\
Sinks and effective sinks			 & 450 & 598 & 286 & 198 & 369 & 275 & 204 & 333 \\
Av. outdegree								 & 1.67 & 1.84 & 4.17 & 3.36 & 4.71 & 4.07 & 3.28 & 4.58 \\
Max outdegree								 & 52 & 51 & 155 & 143 & 336 & 138 & 132 & 292 \\
Max indegree								 & 8 & 10 & 40 & 29 & 49 & 35 & 27 & 44 \\
Self-loops							 & 27 & 31 & 0 & 0 & 0 & 0 & 0 & 0 \\
Undirected edges					 & 0 & 4 & 1238 & 738 & 1468 & 1000 & 596 & 1214 \\
Diameter											 & 9 & 9 & 12 & 15 & 12 & 13 & 14 & 12 \\
Max cycle cluster size					 & 6 & 3 & 351 & 280 & 397 & 305 & 199 & 337 \\
Av. clustering coeff				 & 0.0348 & 0.0421 & 0.1878 & 0.1973 & 0.2446 & 0.1751 & 0.1935 & 0.2389
\end{tabular}
\caption{{\bf Properties of the largest weakly connected differential subnetworks for all cell types.} I = IMR-90 (normal), A = A549 (cancer), H = NCI-H358 (cancer), N = Na\"{\i}ve (normal), M = Memory (normal), D = DLBCL (cancer), F = Follicular lymphoma (cancer), L = EBV-immortalized lymphoblastoma (cancer).\label{table:model_2}}
\end{table*}


\begin{table*}[!ht]
  \begin{tabular}{ c | l c | l c | l c | l c }

	\multirow{3}{*}{ } & \multicolumn{4}{c|}{I/A} & \multicolumn{4}{c}{I/H} \\ 

	& \multicolumn{2}{c}{$p=1$} & \multicolumn{2}{c|}{$p=2$} & \multicolumn{2}{c}{$p=1$} & \multicolumn{2}{c}{$p=2$}  \\ 

	& Gene & $I$ & Gene & $I$ & Gene & $I$ & Gene & $I$ \\ \hline

	\multirow{5}{*}[-0.33cm]{UNC} 
 & HNF1A & 29 & OR5I1 & 35 & HNF1A & 29 & HMX1 & 41 \\
 & TMEM37 & 22 & TMEM37 & 25 & MAP3K3 & 18 & PBX1 & 38 \\
 & OR5I1 & 20 & HNF1A & 23 & TP53 & 18 & MYB & 25 \\
 & MAP3K14 & 19 & POSTN & 21 & RUNX1 & 17 & ITGB2 & 20 \\
 & MAP3K3 & 18 & RORA & 18 & RORA & 16 & TNFRSF10A & 18 \\ \hline

	\multirow{2}{*}{CON} 
 & MAP3K14 & 19 & SRC & 15 & TTN & 16 & BMPR1B & 18 \\
 & SRC & 14 & BMPR1B & 7 & RIPK3 & 6 & LCK & 8 \\

\end{tabular}
\vspace{.3 cm}

  \begin{tabular}{ c | l c | l c | l c | l c | l c | l c }

	\multirow{3}{*}{ } & \multicolumn{4}{c|}{N/D} & \multicolumn{4}{c|}{N/F} & \multicolumn{4}{c}{N/L} \\ 

	& \multicolumn{2}{c}{$p=1$} & \multicolumn{2}{c|}{$p=2$} & \multicolumn{2}{c}{$p=1$} & \multicolumn{2}{c|}{$p=2$} & \multicolumn{2}{c}{$p=1$} & \multicolumn{2}{c}{$p=2$} \\ 

	& Gene & $I$ & Gene & $I$ & Gene & $I$ & Gene & $I$ & Gene & $I$ & Gene & $I$ \\ \hline

	\multirow{5}{*}[-0.2cm]{UNC} 
 & BCL6 & 12 & NFIC & 22 & BCL6 & 12 & NCOA1 & 20 & RBL2 & 11 & RBL2 & 22 \\
 & MEF2A & 5 & TGIF1 & 19 & MEF2A & 5 & NFATC3 & 15 & FOXM1 & 8 & ATF2 & 12 \\
 & NCOA1 & 5 & BCL6 & 14 & NCOA1 & 5 & BCL6 & 11 & ATF2 & 7 & NFATC3 & 11 \\
 & TGIF1 & 4 & FOXJ2 & 12 & TGIF1 & 4 & CEBPD & 8 & RXRA & 5 & RXRA & 9 \\
 & NFATC3 & 4 & NFATC3 & 12 & NFATC3 & 4 & RELA & 8 & NFATC3 & 4 & PATZ1 & 8 \\ \hline

	\multirow{2}{*}{CON} 
 & BUB1B & 2 & CSNK2A2 & 2 & BUB1B & 2 & WEE1 & 2 & BUB1B & 2 & PRKCD & 2 \\
 & AAK1 & 1 & AKT1 & 2 & AAK1 & 1 & CSNK2A2 & 2 & AAK1 & 1 & AURKB & 2 \\
\end{tabular}

\vspace{.3cm}

\begin{tabular}{ c | l c | l c | l c | l c | l c | l c }

	\multirow{3}{*}{ } & \multicolumn{4}{c|}{M/D} & \multicolumn{4}{c|}{M/F} & \multicolumn{4}{c}{M/L} \\ 

	& \multicolumn{2}{c}{$p=1$} & \multicolumn{2}{c|}{$p=2$} & \multicolumn{2}{c}{$p=1$} & \multicolumn{2}{c|}{$p=2$} & \multicolumn{2}{c}{$p=1$} & \multicolumn{2}{c}{$p=2$} \\
	& Gene & $I$ & Gene & $I$ & Gene & $I$ & Gene & $I$ & Gene & $I$ & Gene & $I$ \\ \hline

	\multirow{5}{*}[-0.2cm]{UNC} 
 & BCL6 & 12 & FOXJ2 & 12 & BCL6 & 12 & NCOA1 & 18 & RBL2 & 11 & RBL2 & 16 \\
 & MEF2A & 5 & NFIC & 12 & MEF2A & 5 & BCL6 & 13 & FOXM1 & 8 & ATF2 & 10 \\
 & NCOA1 & 5 & BCL6 & 11 & NCOA1 & 5 & E2F3 & 9 & ATF2 & 7 & ZNF91 & 8 \\
 & NFATC3 & 4 & NCOA1 & 9 & NFATC3 & 4 & RUNX1 & 9 & RXRA & 5 & STAT6 & 8 \\
 & SMAD4 & 4 & MEF2A & 8 & RELA & 4 & TFE3 & 7 & TGIF1 & 4 & FOXM1 & 8 \\ \hline

	\multirow{2}{*}{CON} 
 & AAK1 & 1 & RIPK2 & 1 & AAK1 & 1 & ROCK2 & 2 & AAK1 & 1 & AURKB & 2 \\
 & RIPK2 & 1 & MAST2 & 1 & RIPK2 & 1 & RIPK2 & 1 & SCYL3 & 1 & RIPK2 & 1 \\
\end{tabular}
\caption{{\bf Best single genes and their impacts for the $p$=1 and $p$=2 models.} The unconstrained (UNC) and constrained (CON) case are shown. The constrained case refer to target that are kinases and are expressed in the cancer case. I = IMR-90 (normal), A = A549 (cancer), H = NCI-H358 (cancer), N = Na\"{\i}ve (normal), M = Memory (normal), D = DLBCL (cancer), F = Follicular lymphoma (cancer), L = EBV-immortalized lymphoblastoma (cancer). \label{table:best_singles}}
\end{table*}




\newpage

\section*{Figure Legends}



\end{document}